\documentclass[]{aa}
\usepackage{sidecap}
\usepackage{graphicx}
\usepackage[varg]{txfonts}
\usepackage{url}

\usepackage{natbib}
\bibpunct{(}{)}{;}{a}{}{,}
\usepackage{lscape}
\usepackage[dvipsnames]{xcolor}
\begin{document} 
\title{Reconstructing the mid-infrared environment in the stellar merger remnant V838 Monocerotis}
   \author{Muhammad Zain Mobeen \inst{1}
          \and
          Tomasz Kami{\'n}ski\inst{1}
          \and
          Alexis Matter\inst{2}
          \and
          Markus Wittkowski\inst{3}
          \and 
          John D. Monnier\inst{5}
          \and
          Stefan Kraus\inst{6}
          \and
          Jean-Baptiste Le Bouquin\inst{7}          
          \and
          Narsireddy Anugu\inst{8}
          \and
          Theo Ten Brummelaar\inst{8}  
          \and
          Claire L. Davies\inst{6}
          \and
          Jacob Ennis\inst{5}
          \and
          Tyler Gardner\inst{6}
          \and
          Aaron Labdon\inst{4} 
          \and 
          Cyprien Lanthermann\inst{8} 
          \and 
          Gail H. Schaefer\inst{8} 
          \and 
          Benjamin R. Setterholm\inst{5}
          \and
          Nour Ibrahim\inst{5}
          \and
          Steve B. Howell\inst{9}
}

\institute{\centering
Nicolaus Copernicus Astronomical Center, Polish Academy of Sciences, Rabia{\'n}ska 8, 87-100 Toru{\'n}, Poland %\email{mzainmob@ncac.torun.pl}\label{inst1}
%\and
\and
Université Côte d’Azur, Observatoire de la Côte d’Azur, CNRS, Laboratoire Lagrange, France  \label{inst2}
\and
European Southern Observatory, Karl-Schwarzschild-Str. 2, 85748 Garching bei Munchen, Germany\label{inst3}
\and 
European Southern Observatory, Alonso de Cordoba 3107, Vitacura, Santiago, Chile\label{inst4}
\and
Astronomy Department, University of Michigan, Ann Arbor, MI 48109, USA\label{inst5}
\and
Astrophysics Group, Department of Physics \& Astronomy, University of Exeter, Stocker Road, Exeter, EX4 4QL, UK\label{inst6}
\and
Institut de Planetologie et d'Astrophysique de Grenoble, Grenoble 38058, France\label{inst7}
\and
The CHARA Array of Georgia State University, Mount Wilson Observatory, Mount Wilson, CA 91203, USA\label{inst8}
\and
NASA Ames Research Center, Moffett Field, CA 94035, USA \label{inst9}
}

%\date{Received March, 2020}
\authorrunning{Mobeen et al.}
\titlerunning{MATISSE observations of V838 Mon}
% \abstract{}{}{}{}{} 
% 5 {} token are mandatory
 
\abstract{V838 Mon is a stellar merger remnant that erupted in 2002 in a luminous red novae event. Although it is well studied in the optical, near infrared and submillimeter regimes, its structure in the mid infrared wavelengths remains elusive. Over the past two decades only a handful of mid-infrared interferometric studies have been carried out and they suggest the presence of an elongated structure at multiple wavelengths. However, given the limited nature of these observations, the true morphology of the source could not be determined.}{By performing image reconstruction using observations taken at the VLTI and CHARA, we aim to map out the circumstellar environment in V838 Mon.}{We observed V838 Mon with the MATISSE ($LMN$ bands) and GRAVITY ($K$ band) instruments at the VLTI and also the MIRCX/MYSTIC ($HK$ bands) instruments at the CHARA array. We geometrically modelled the squared visibilities and the closure phases in each of the bands to obtain constraints on physical parameters. Furthermore, we constructed high resolution images of V838 Mon in the $HK$ bands using the MIRA and SQUEEZE algorithms to study the immediate surroundings of the star. Lastly we also modelled the spectral features seen in the $K$ and $M$ bands at various temperatures.}{The image reconstructions show a bipolar structure that surrounds the central star in the post merger remnant. In the $K$ band the super resolved images show an extended structure (uniform disk diameter $\sim 1.94$ mas) with a clumpy morphology that is aligned along a north-west position angle (PA) of --$40\degr$. Whereas in the $H$ band the extended structure (uniform disk diameter $\sim 1.18$ mas) lies roughly along the same PA. However the northern lobe is slightly misaligned with respect to the southern lobe which results in the closure phase deviations.}{The VLTI and CHARA imaging results show that V838 Mon is surrounded by features that resemble jets that are intrinsically asymmetric. This is further confirmed by the closure phase modelling. Further observations with VLTI can help to determine whether this structure shows any variation over time and also if such bi-polar structures are commonly formed in other stellar merger remnants. }
%\vspace{2cm}

\maketitle

\section{Introduction}\label{intro}

 At the start of 2002 V838 Monocerotis erupted in a luminous red nova event \citep{2002A&A...389L..51M,2005A&A...436.1009T} and in a few weeks brightened by almost two orders of magnitude, finally reaching a peak luminosity of $10^{6} L_{\sun}$ \citep{2005A&A...436.1009T,2008AJ....135..605S,2003Natur.422..405B}. The event is thought to have been the result of a stellar merger. According to the scenario proposed in \cite{2006A&A...451..223T}, an 8 $M_{\sun}$ B-type main sequence star coalesced with a 0.4 $M_{\sun}$ young stellar object. The outburst was soon followed by a gradual decrease in temperature, and its spectra soon evolved to that of a late M-type supergiant \citep{2003MNRAS.343.1054E, 2015AJ....149...17L}. Spectra taken in the 2000s revealed the presence of various molecules in V838 Mon, including water and transition-metal oxides \citep{ref161B,2009ApJS..182...33K}. Dust was also observed to be produced in the post merger environment \citep{2008ApJ...683L.171W,alma}. Additionally, a B-type companion was observed in the vicinity of the central merger remnant, which thus suggests that the merger had taken place in a hierarchical triple system \citep{Bdiscovery2,alma}. The companion was obscured by dust formed in the aftermath of the 2002 eruption \citep{2009A&A...503..899T}. V838 Mon is the best studied luminous red nova in the Milky Way, although, many others have also been found within the Galaxy as well as elsewhere in the Local Group \citep{2019A&A...630A..75P}. 
 
 As the merger remnant in V838 Mon is enshrouded by dust, it is therefore an ideal target for mid-infrared interferometric studies. The first of these studies was conducted by \cite{2005ApJ...622L.137L} in which they observed V838 Mon using the Palomar Testbed Interferometer (PTI). By modelling the squared visibilities in the $K$-band at 2.2\,$\mu$m they were able to measure the size of the merger remnant of $1.83 \pm 0.06$ mas. There were also hints of asymmetries in the object, but due to scarce measurements these could not be confirmed. \cite{2014A&A...569L...3C} followed up these measurements between 2011 and 2014, using the Very Large Telescope Interferometer (VLTI) instruments: Astronomical Multi-BEam combineR \citep[AMBER;][]{2007A&A...464....1P} in $H$ and $K$ bands, and the MID-infrared Interferometric instrument \citep[MIDI;][]{2003Ap&SS.286...73L} in $N$ band. Fitting uniform disk models to the AMBER measurements gave an angular diameter of $1.15 \pm 0.2$ mas, which --according to the authors -- indicates that the photosphere in V838 Mon had contracted by about 40\% over the course of a decade. Also, their modeling of the AMBER data suggests that an extended component was present in the system, with a lower limit on the full width at half-maximum (FWHM) of $\sim$20\,mas. Modeling of the MIDI measurements seems to point towards the presence of a dusty elongated structure whose major axis varies as a function of wavelength between 25 and 70\,mas in $N$ band. Submillimeter observations obtained with the Atacama Large Millimeter/sub millimeter Array (ALMA) in continuum revealed the presence of a flattened structure with a FWHM of 17.6$\times$17.6 mas surrounding V838 Mon \citep{alma}. Recent $L$ band measurements by \cite{2021A&A...655A.100M} also seem to paint a similar picture. \cite{2021A&A...655A.100M} geometrically modelled the squared visibilities and closure phases in the $L$-band, obtained using the Multi AperTure mid-Infrared SpectroScopic Experiment instrument (MATISSE) at the VLTI in 2020. They found that the structure in the $L$-band is well represented by an elliptical disk tilted at an angle of --40$\degr$. Furthermore, the closure phases showed small but non-zero deviations, which suggest the presence of asymmetries in the system. The interferometric measurements span across the wavebands (from 2.2 $\mu$m to 1.3 mm) and trace a dusty structure oriented roughly along the same direction, with PA in the range --10\degr\ (MIDI) to --50\degr\  (ALMA). This might indicate either a single overarching structure in the post-merger remnant, or multiple similarly aligned structures. Simulations of stellar merger events also suggest the presence of a disk like structure in the post-merger remnant, which is thought to be a reservoir for the pre-merger binary angular momentum \citep[e.g.][]{1976ApJ...209..829W, 2017ApJ...850...59P}. 
 
 V838 Mon serves as an excellent source to advance our understanding of the post-merger environment decades after the luminous red nova event. Thus, it provides us with crucial insights into the physical processes at play in these merger events and their final products in the long term. In this paper, we analyze and interpret recent interferometric observations obtained with a variety of instruments that span many near to mid infrared wavelengths.       

 The format of the paper is as follows. In Sect. \ref{Observations} we present all of our VLTI and CHARA observations and outline the main steps of the data reduction. We also analyze and interpret recent optical speckle interferometric observations obtained at 562 nm and 832 nm. In Sect. \ref{geomod} we mainly present the results of geometrically modelling the interferometric observables (squared visibilities and closure phases) observed with the MATISSE and GRAVITY instruments at VLTI and with MIRCX/MYSTIC at CHARA. Sect \ref{imaging} centers around our image reconstruction attempts for the VLTI and CHARA datasets using two distinct image reconstruction algorithms. The modelling and imaging results are discussed in depth in Sect. \ref{disc} finally followed by Sect. \ref{concl} in which we present the main conclusions of this study.     
%%%%%%%%%%%%%%%%%%%%%%%%%%%%%%%%%%%%%%%%%%%%%%%%%%%%%%%%%%%%%%%%%%%%%%%%%%%%%%%%%%%%%%%%%%%%%%%%%%%%%%%%%%%%%%%%%%%%%%%%%%%%%%%%%%%%
\section{Observations and data reduction}\label{Observations}

\begin{figure}%[hbt!]
    \centering
    \includegraphics[clip, width=\columnwidth]{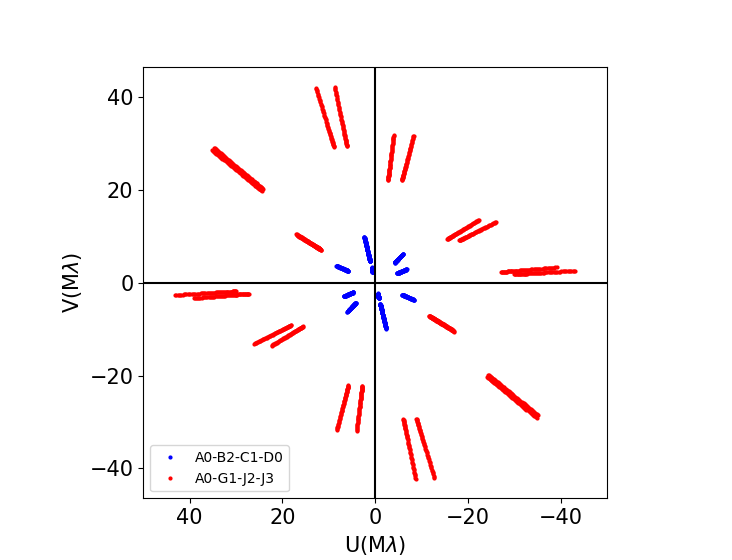}
    \caption{UV tracks for MATISSE observing runs in 2021--2022.}
    \label{Fig-MATISSEtracks}
\end{figure}

\begin{figure}%[hbt!]
    \centering
    \includegraphics[clip, width=\columnwidth]{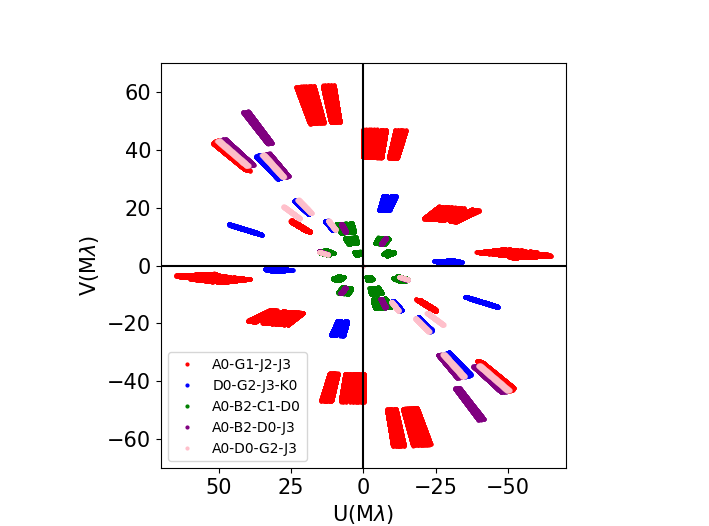}
    \caption{UV coverage for GRAVITY observing runs.}
    \label{Fig-GRAVITYtracks}
\end{figure}

V838 Mon was observed with the VLTI located at Paranal observatory in Chile, in 2021 and 2022. Observations were carried out using the 1.8 m Auxiliary Telescopes (ATs) and two instruments, MATISSE \citep{2022A&A...659A.192L} and GRAVITY \citep{2017A&A...602A..94G}. MATISSE is a four telescope beam combiner which covers the $L$ (2.8--4.2 $\mu$m), $M$ (4.5--5 $\mu$m) and $N$ (8--13 $\mu$m) bands, while GRAVITY combines light in the $K$ (1.9--2.4 $\mu$m) band. For both instruments, we intended to get 18 observing blocks (OBs) to perform image reconstruction. However, we were only able to obtain three OBs for MATISSE and fifteen OBs for GRAVITY. The technical details of our VLTI observations are presented in Tables \ref{gravity_log}--\ref{matisse_log}. In the case of MATISSE, only the large and small configurations were employed, while for GRAVITY, observations using all three configurations were carried out. In particular, intermediate configurations were also used to sample better the UV plane. The UV coverages obtained for the MATISSE and GRAVITY observations are shown in Figs. \ref{Fig-MATISSEtracks} \& \ref{Fig-GRAVITYtracks}, respectively.  

MATISSE observations were carried out in low spectral resolution (R$\sim 30$) using the GRA4MAT mode, in which the GRAVITY fringe tracker is used to stabilize fringes for MATISSE \citep{2022A&A...659A.192L}. Each MATISSE observation for V838 Mon consisted of a CAL-SCI-CAL observing sequence in which two calibrator stars, 20 Mon (spectral type K0 III) and HD 52666 (spectral type K5 III) were observed to calibrate the $LMN$ bands. The source and calibrator fluxes in the $LMN$ bands are given in Table \ref{table:Table-4}.  

The GRAVITY observations were carried out in medium spectral resolution (R$\sim$500) using its single-field mode in which the light is split equally between the fringe tracker channel and the science channel. We adopted a CAL-SCI-CAL sequence for each GRAVITY run. However, one calibrator was observed here, HD 54990 (spectral type K0/1 III).    

Interferometric observations of V838 Mon were also obtained at the Center for High Angular Resolution Astronomy (CHARA) Array on UT 2022 Mar 2, 3, and 9. The CHARA Array is operated by Georgia State University and is located at Mount Wilson Observatory in southern California. The array consists of six 1-meter telescopes arranged in a Y-configuration with baselines ranging in length from 34 m to 331 m \citep{2005ApJ...628..453T}. We combined the light from 4 to 5 of the telescopes using the Michigan InfraRed Combiner-eXeter (MIRC-X) beam combiner in the $H$-band using the low spectral resolution (R=50) prism that disperses light over 8 spectral channels \citep{2020AJ....160..158A}. On the last night, additional simultaneous observations were obtained using the Michigan Young STar Imager (MYSTIC) combiner \citep{2022SPIE12183E..0BS} in the $K$-band in low spectral resolution (R=49) mode \citep{2018SPIE10701E..22M}. The CHARA MIRC-X and MYSTIC observation log is given in Table \ref{chara log}. On each night, we alternated between observations of V838 Mon and calibrator stars to calibrate the interferometric transfer function. The $H$ band fluxes are presented in Table \ref{table:Table-6}, while the uniform disk diameters of the calibrators in the $HK$ bands were adopted from \cite{2014ASPC..485..223B} and \cite{2017yCat.2346....0B} and are listed in Table \ref{table:Table-7}. 

\begin{figure}%[hbt!]
    \centering
    \includegraphics[clip, width=\columnwidth]{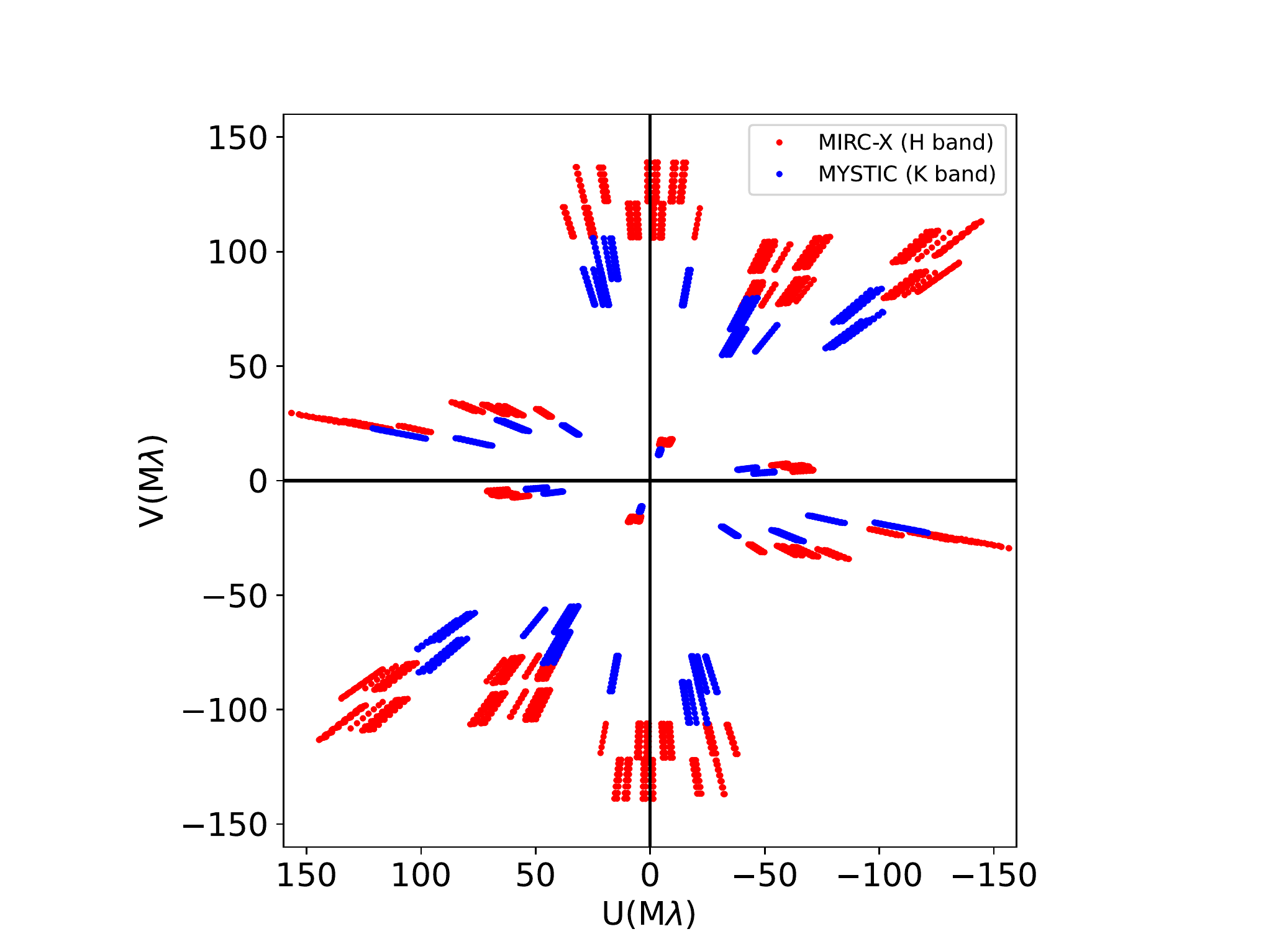}
    \caption{UV coverage for CHARA observing runs in the $HK$ bands.}
    \label{Fig-CHARAtracks}
\end{figure}

\subsection{MATISSE data reduction}
The $LMN$-band MATISSE data were reduced using version 1.7.5 of the ESO data reduction pipeline. The final products were OIFITS files (version 2) comprising uncalibrated interferometric observables, including six dispersed squared visibilities and three independent closure phase measurements per exposure. The observations of V838\,Mon were then calibrated using our measurements of the calibrator stars listed in Table \ref{table:Table-4}, whose diameters and fluxes were taken from the Mid-infrared stellar Diameters and Fluxes compilation Catalog\footnote{MDFC is available through VizieR service at https://vizier.u-strasbg.fr/viz-bin/VizieR} \citep[MDFC;][see Table\,\ref{matisse_log}]{2019yCat.2361....0C}. The calibration of squared visibilities was performed by dividing the raw squared visibilities of V838 Mon by those of the calibrators, corrected for their diameter (which is called the interferometric transfer function). The calibrated $LMN$ spectra of V838\,Mon were obtained by multiplying the ratio between the target and calibrator raw fluxes measured by MATISSE at each wavelength by a model of the absolute flux of the calibrator. This model was taken from the PHOENIX stellar spectra grid \citep{2013A&A...553A...6H}. 

%In order to remove any remaining telluric absorption features from the observed $M$ band spectra, we continuum normalized it using spectra of various giant stars found in the Cassini Atlas of Stellar Spectra \citep{2015ApJS..221...30S}. This allowed us to compare the observed spectra with the modeled synthetic spectra. 

Upon completing the data reduction and inspecting the resultant products, we found that the $N$ band squared visibilities were of poor quality. This was due to the fact that the average $N$-band total flux of V838 Mon ($\sim$15\,Jy) lies very close to the photometric sensitivity limit of MATISSE in the $N$ band. Furthermore, due to the increased thermal background effects beyond 11\,$\mu$m, the quality of the photometric data became even worse. This rendered the $N$-band photometry, and thus the absolute visibilities, unusable, and they were discarded. However, we were able to make use of the correlated fluxes ($\sim$5\,Jy), which turn out to be slightly greater than the sensitivity limit of the GRA4MAT mode in $N$-band. Those observables allowed us to model geometrically the source in the $N$ band and put constraints on the extent of the structure in this band.

\subsection{GRAVITY data reduction}
GRAVITY data reduction was performed with the ESO GRAVITY data reduction pipeline version 1.4.1 This was carried out in the ESO Reflex workflow environment. For the calibration of squared visibilities we used the visibility calibration workflow in Reflex. Furthermore, the reduced $K$ band spectra were flux calibrated similarly to the $L$ band spectra obtained from MATISSE. The flux calibration routine from the MATISSE consortium was used for this purpose and, just like in the case of the $M$ band, the $K$ band spectra were also similarly normalized and later compared to synthetic spectra.    

\subsection{CHARA data reduction}

CHARA data was reduced and calibrated by the support astronomers at the CHARA array. The data were reduced using the standard MIRC-X and MYSTIC pipeline (version 1.3.5) written in python\footnote{https://gitlab.chara.gsu.edu/lebouquj/mircx\_pipeline.git} and described by \cite{2020AJ....160..158A}. The pipeline produces calibrated visibility amplitudes for each pair of telescopes and closure phases for each combination of three telescopes. To assess the data quality, the calibrators were checked against each other on each night. The calibrators showed no evidence for binarity based on a visual inspection of the data, and the diameters derived from the measured visibilities were consistent with the expected values within uncertainties. The calibrated OIFITS files will be available in the Optical Interferometry Database\footnote{https://oidb.jmmc.fr/index.html} and the CHARA Data Archive\footnote{https://www.chara.gsu.edu/observers/database}.

\subsection{Gemini South observation}

V838 Mon was also observed twice with the Gemini South 8-m telescope using the high-resolution Zorro instrument \citep{2021FrASS...8..138S,2022FrASS...9.1163H}. Zorro provides simultaneous speckle imaging in two optical bands, here centered at 562 nm and 832 nm. V838 Mon was observed on 25 Feb 2021 UT and, about one year later, on 20 March 2022 UT. The speckle imaging on each night consisted of a number of 60 ms frames taken in a row, the February 2021 observation taking 12,000 frames and the March 2022 observation taking 16,000 frames. The February observations occurred during a night of average seeing (0\farcs6) while the longer March 2022 observations occurred during good seeing (0\farcs45). The data were reduced using a  standard speckle imaging pipeline with reduced output data products including reconstructed images and 5$\sigma$ contrast limits \citep{2011AJ....142...19H}. The two sets of observations agreed well with the March 2022 data, providing a higher S/N result. 

\begin{figure}%[hbt!]
    \centering
    \includegraphics[clip, width=\columnwidth]{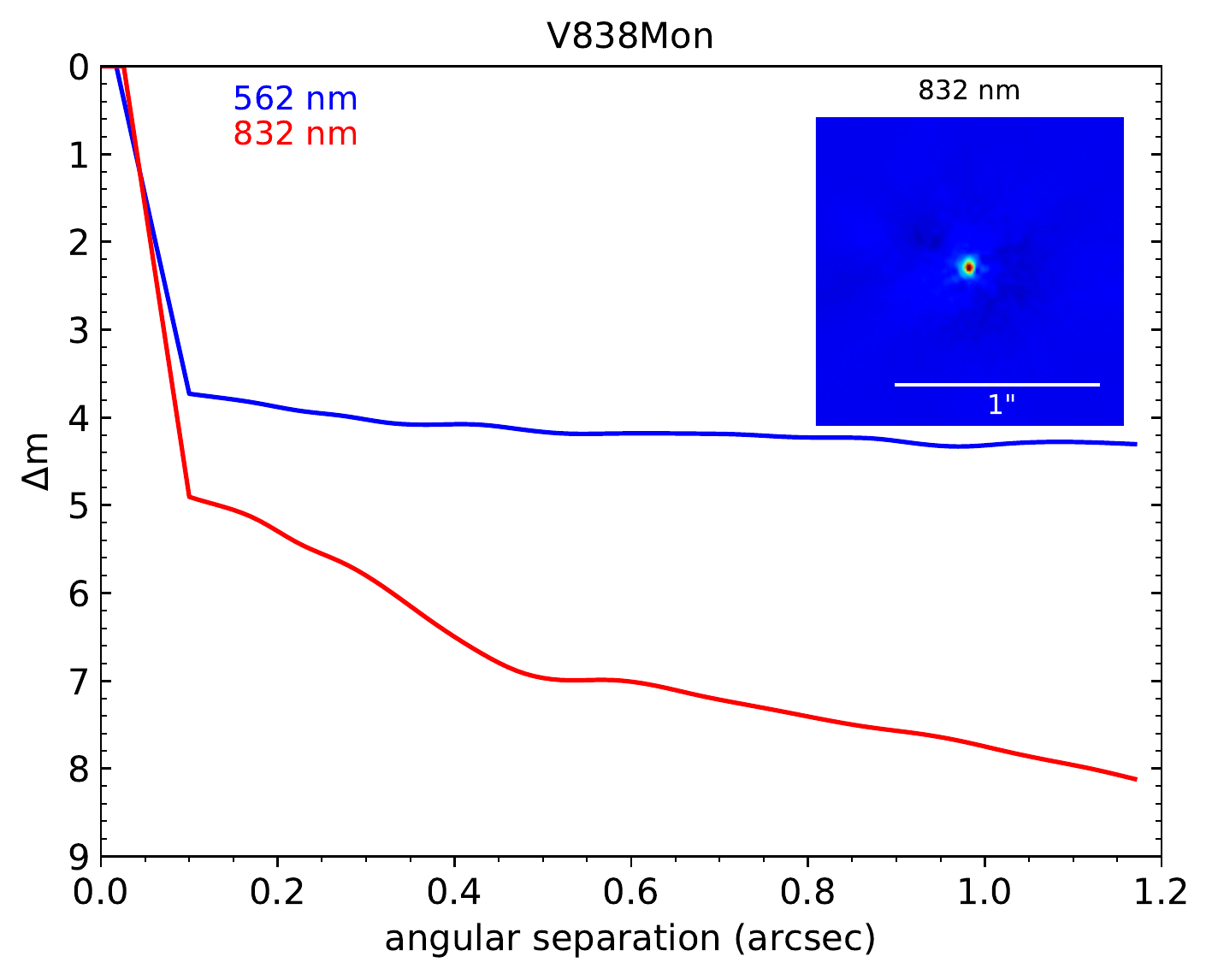}
    \caption{Contrast (in magnitudes) curves for both filters from March 2022. The speckle image reconstruction at 832 nm is shown in the top right corner.}
    \label{gemini-contrast-2022}
\end{figure}

\begin{figure}%[hbt!]
    \centering
    \includegraphics[clip, width=\columnwidth]{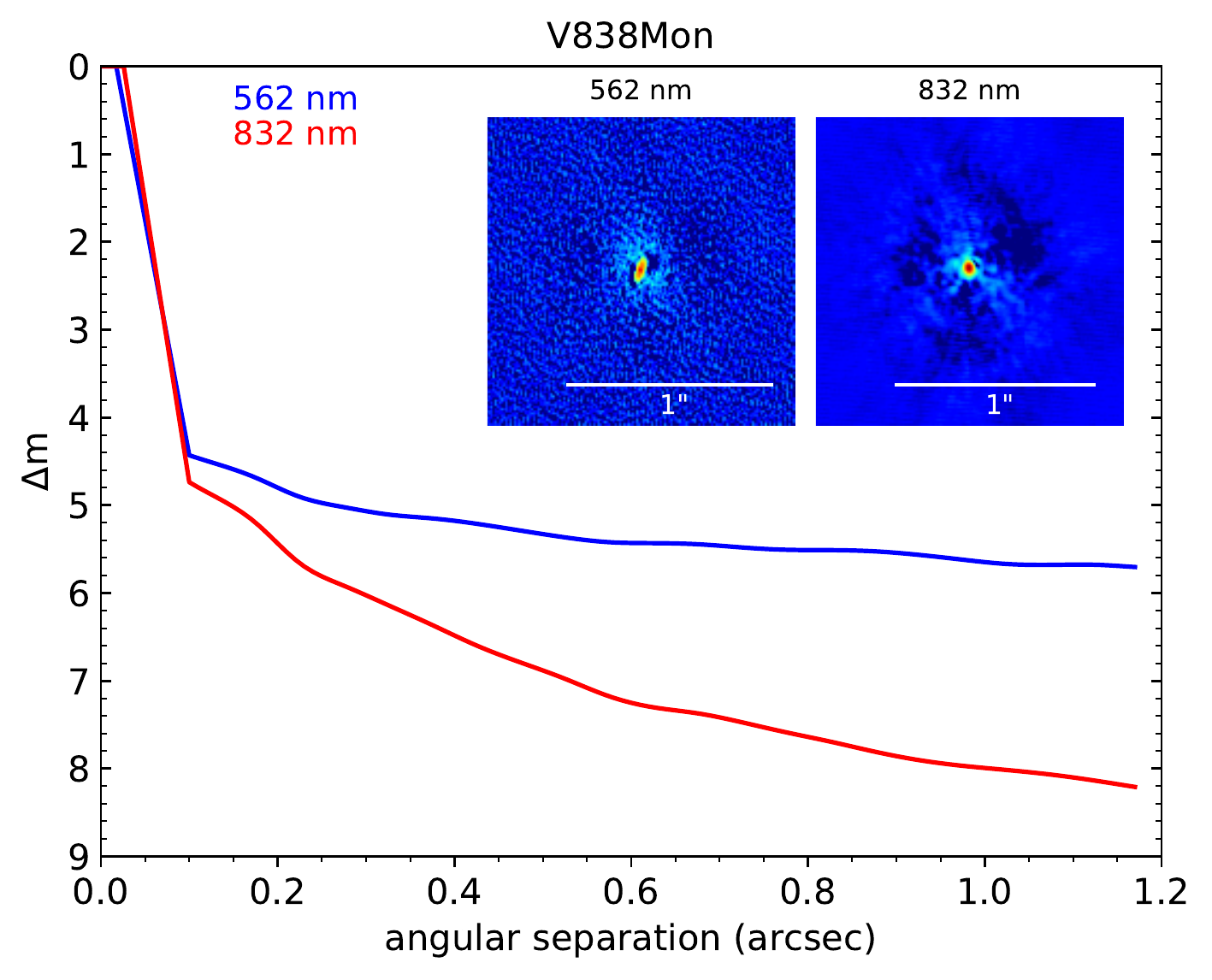}
    \caption{Same as \ref{gemini-contrast-2021} but for February 2021. The speckle image reconstruction at 562 nm is also shown.}
    \label{gemini-contrast-2021}
\end{figure}

\begin{figure}%[hbt!]
    \centering
    \includegraphics[clip, width=\columnwidth]{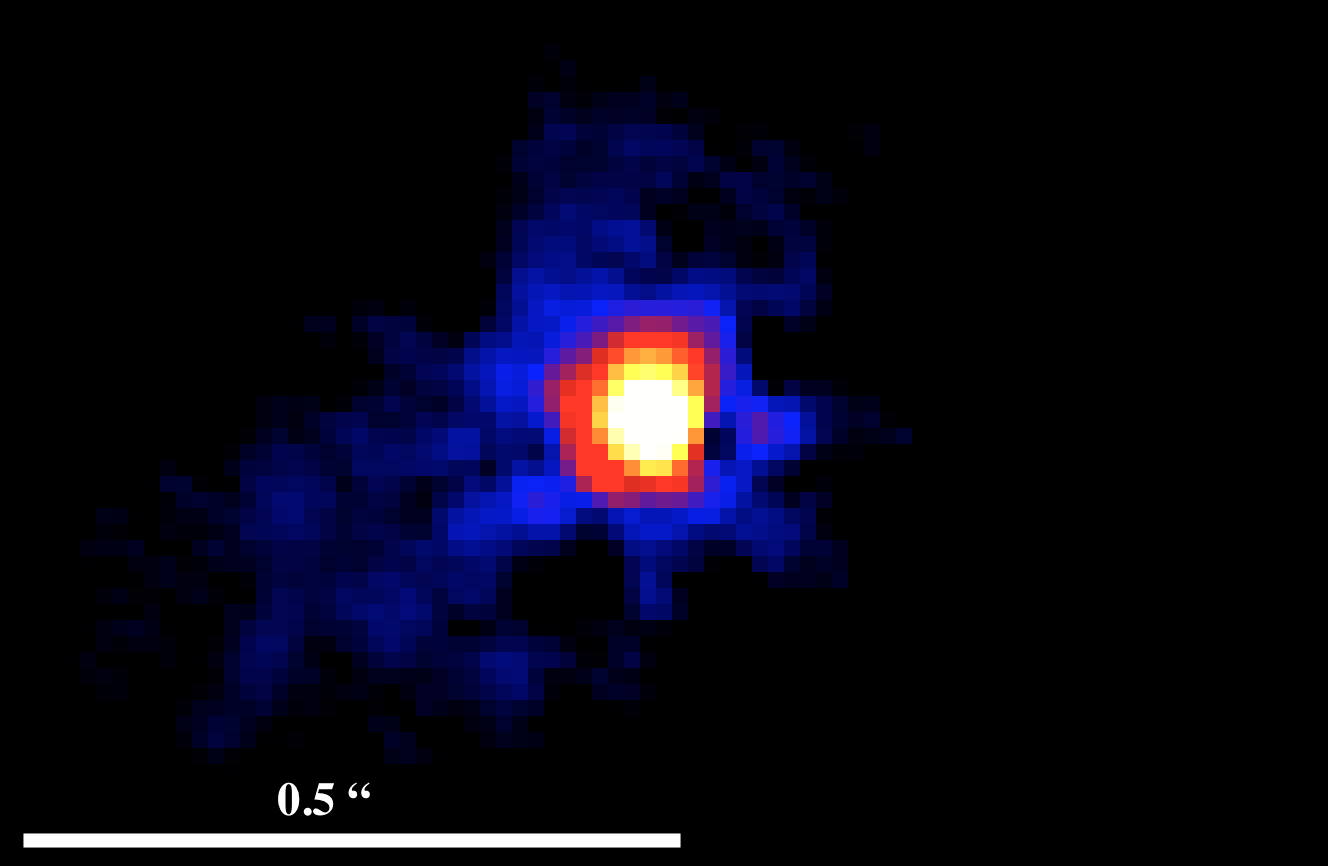}
    \caption{Speckle image reconstruction from March 2022 at 832 nm. A noticeable north-south elongation is visible. The image field of view (FoV) is 0\farcs85 by 0\farcs55. North is up, east is left.}
    \label{gemini-Iband}
\end{figure}

\begin{figure}%[hbt!]
    \centering
    \includegraphics[clip, width=\columnwidth]{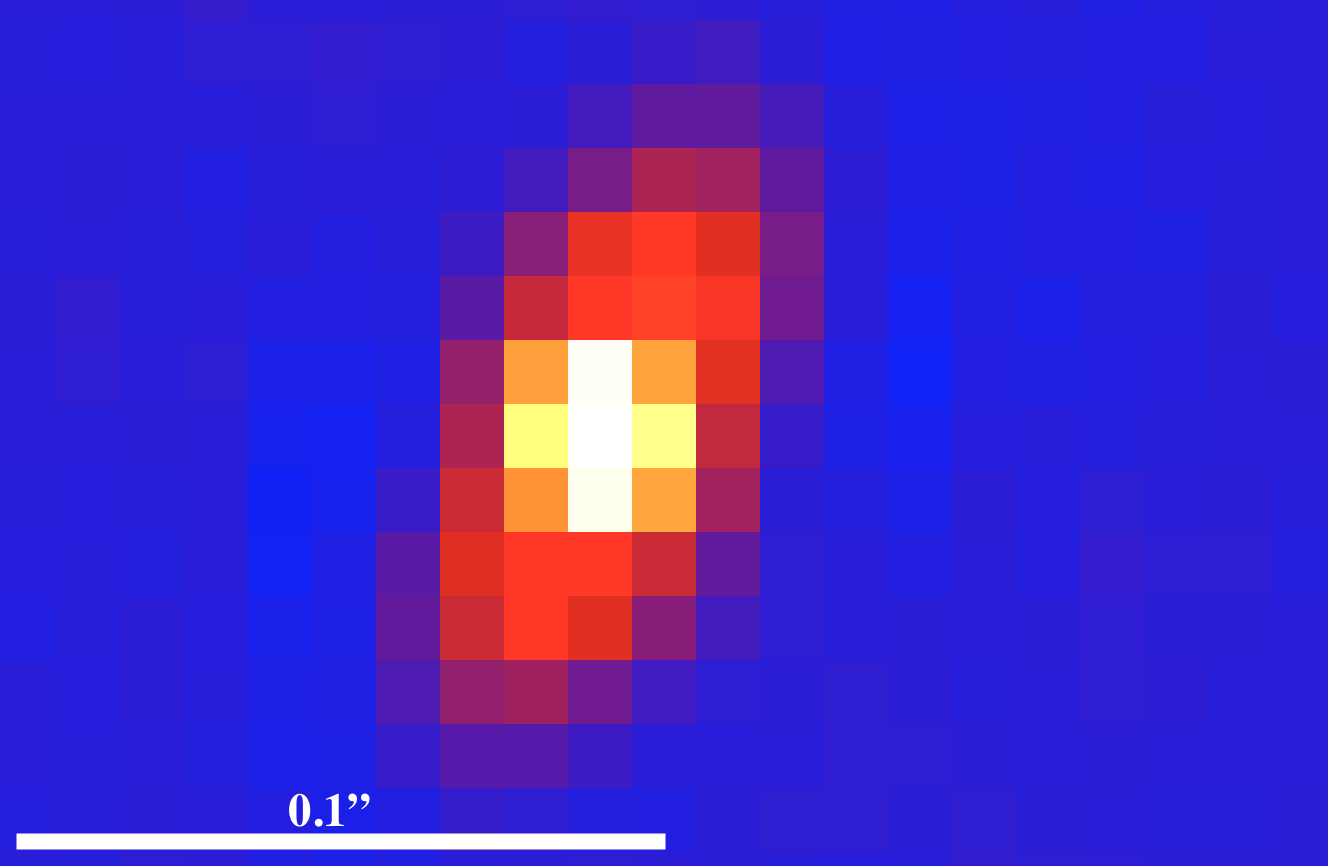}
    \caption{Speckle image reconstruction from February 2021 at 562 nm. A prominent north-west elongation is visible. The image FoV is 0\farcs21 by 0\farcs14. North is up, east is left.}
    \label{gemini-vband}
\end{figure}

Fig. \ref{gemini-contrast-2022} shows the resultant contrast curves for both filters and our 832\,nm reconstructed image from March 2022. Fig. \ref{gemini-contrast-2021} shows a similar plot but for February 2021 along with image reconstructions at 832 nm and at 562 nm. There are no close (<1\farcs2) stellar companions detected within the angular and contrast limits achieved. However, as seen in detail in Fig.\,\ref{gemini-Iband}, the image at 832 nm is extended beyond just a point source with a slight elongation in the north-south direction and to the east by about 0\farcs03. Similarly, at 562\,nm, we see also see a noticeable elongation in the north-western direction, as shown in Fig.\,\ref{gemini-vband}. The 562\,nm elongation is in very good agreement with what we observe in the mid-infrared bands (see Fig.\,\ref{fig-cartoon}). While the elongation varies significantly at 832\,nm, we note that the orientation (north-south) of this particular structure is remarkably similar to the PA ($\sim$3\degr) we obtain by fitting the $N$ band visibility amplitudes with an elliptical disk model. Only with future extensive observations at Gemini South will it be possible to reliably constrain the orientation of the elongation of V838\,Mon in the $V$ and $I$ bands.

\begin{table}%[ht]
\caption{VLTI/GRAVITY observation log.}
\centering
\begin{tabular}{c c c}
\hline
Date & Configuration & Seeing  \\ % inserts table %heading                                             \\
\hline
01/04/2022&A0-G1-J2-J3&0\farcs42 \\
02/03/2022&A0-B2-C1-D0&0\farcs86  \\
02/03/2022&A0-B2-C1-D0&0\farcs82 \\
02/04/2022&A0-B2-C1-D0&0\farcs90 \\
02/04/2022&A0-B2-C1-D0&1\farcs00 \\
02/04/2022&A0-B2-C1-D0&0\farcs62 \\
02/05/2022&A0-D0-G2-J3&0\farcs57 \\
02/06/2022&D0-G2-J3-K0&0\farcs50 \\ 
02/06/2022&D0-G2-J3-K0&1\farcs04 \\
02/10/2022&D0-G2-J3-K0&0\farcs56 \\
02/11/2022&A0-G1-J2-J3&0\farcs69 \\
02/13/2022&A0-G1-J2-J3&0\farcs66 \\ 
03/11/2022&A0-G1-J2-J3&0\farcs55 \\
03/25/2022&A0-G1-J2-J3&0\farcs41 \\
03/25/2022&A0-G1-J2-J3&0\farcs40\\

\hline
\end{tabular}
\label{gravity_log}
\end{table}

\begin{table}%[ht]
\caption{VLTI/MATISSE observation log.}
\centering
\begin{tabular}{c c c}
\hline
Date & Configuration & Seeing \\ % inserts table %heading                                             \\
\hline
10/30/2021&A0-G1-J2-J3&0\farcs41 \\
03/02/2022&A0-G1-J2-J3&0\farcs45  \\
03/31/2022&A0-B2-C1-D0&0\farcs77 \\
\hline
\end{tabular}
\label{matisse_log}
\end{table}

\begin{table}%[ht]
\caption{CHARA observation log.}
\centering
\begin{tabular}{c c c c}
\hline
Date &Combiner &Mode &Configuration \\ % inserts table %heading                                             \\
\hline
03/02/2022&MIRC-X&H-Prism50&S1-S2-E2-W1-W2 \\%&1\farcs24  \\
03/03/2022&MIRC-X&H-Prism50&S1-S2-E2-W1-W2 \\%&1\farcs00  \\
03/09/2022&MIRC-X&H-Prism50&S1-S2-E2-W1-W2 \\%&1\farcs29 \\
03/09/2022&MYSTIC&K-Prism49&S1-S2-E2-W1-W2 \\%&15%\farcs29 \\
\hline
\end{tabular}
\label{chara log}
\end{table}

\begin{table}%[ht]
\caption{Total $LMN$-band fluxes of V838\,Mon and the calibrators.}
%Source and calibrator total fluxes. HD 52666 fluxes were taken from the Mid-infrared stellar Diameters and Fluxes compilation Catalogue (MDFC)  \cite{2019yCat.2361....0C}}
\centering
\begin{tabular}{c c c c c}
\hline
Object & $L$ band & $M$ band & $N$ band & Reference\\  % inserts table %heading
&[Jy]&[Jy]&[Jy]  \\
\hline
V838\,Mon&5&4&30 &1\\
HD\,52666&132&67&18 &2 \\
20\,Mon&30.3&19.2&4.31&2 \\
\hline
\end{tabular}
\tablebib{(1)~\cite{alma}; (2) \cite[MDFC;][]{2019yCat.2361....0C}}
\label{table:Table-4}
\end{table}

\begin{table}%[ht]
\caption{Total $K$-band fluxes of V838\,Mon and the calibrator.}
%Source and calibrator total fluxes. HD 52666 fluxes were taken from the Mid-infrared stellar Diameters and Fluxes compilation Catalogue (MDFC)  \cite{2019yCat.2361....0C}}
\centering
\begin{tabular}{c c c c c}
\hline
Object & $K$ band & Reference\\  % inserts table %heading
&[mag]  \\
\hline
V838\,Mon&5.08 &1\\
HD\,54990&3.77 &2 \\

\hline
\end{tabular}
\tablebib{(1)~\cite{alma}; (2) \cite[MDFC;][]{2019yCat.2361....0C}}
\label{table:Table-5}
\end{table}

\begin{table}%[ht]
\caption{Total $H$-band fluxes of V838\,Mon and the calibrator.}
%Source and calibrator total fluxes. HD 52666 fluxes were taken from the Mid-infrared stellar Diameters and Fluxes compilation Catalogue (MDFC)  \cite{2019yCat.2361....0C}}
\centering
\begin{tabular}{c c c c c}
\hline
Object & $H$ band & Reference\\  % inserts table %heading
&[mag]  \\
\hline
V838\,Mon&5.86 &1\\
HD\,54990&3.89 &2 \\
HD\,61039&5.54 &2\\
HD\,59230&4.16 &2\\

\hline
\end{tabular}
\tablebib{(1)~\cite{alma}; (2) \cite[MDFC;][]{2019yCat.2361....0C}}
\label{table:Table-6}
\end{table}

\begin{table}%[ht]
\caption{$H$-band calibrator uniform disk diameters.}
%Source and calibrator total fluxes. HD 52666 fluxes were taken from the Mid-infrared stellar Diameters and Fluxes compilation Catalogue (MDFC)  \cite{2019yCat.2361....0C}}
\centering
\begin{tabular}{c c c c}
\hline
Calibrator & UDD, $H$ & UDD, $K$ & e$_{\rm UDD}$\\  % inserts table %heading
&[mas] &[mas] &[mas]  \\
\hline
HD\,54990&0.863&0.867&0.079\\
HD\,61039&0.377&0.379&0.009\\
HD\,59230&0.749&0.753&0.072  \\

\hline
\end{tabular}
\label{table:Table-7}
\tablefoot{Uniform disk diameter (UDD) in the $HK$ bands, taken from the JMMC Stellar Diameters Catalog \citep{2014ASPC..485..223B}. e$_{\rm UDD}$ is the errorbar on the diameter measurement.}
\end{table}

%\textbf{As a result of this calibration, the error bars on the squared visibilities increase due to the calibrator noise. Also, some points in the uncalibrated visibilities disappear and are not observed in their calibrated counterparts. This happens due to the poor quality of some calibrator visibility measurements. 

%%%%%%%%%%%%%%%%%%%%%%%%%%%%%%%%%%%%%%%%%%%%%%%%%%%%%%%%%%%%%%%%%%%%%%%%%%%%%%%%%%%%%%%%%%%%%%%%%%%%%%%%%%%%%%%%%%%%%%%%%%%%%%%%%%
\section{Geometrical modelling} \label{geomod}
The resultant data products for the above-mentioned VLTI and CHARA instruments are the interferometric observables, i.e. squared visibilities ($V^{2}$) and the closure phases. The squared visibilities represent the fringe contrast, thus an object is said to be completely resolved in the case that the value for the squared visibility is zero, and completely unresolved in the case the value is one. The squared visibilities can be used to constrain the size of the source. Closure phases are the sum of the individual phase measurements by telescopes within a particular triangular configuration in the array. This results in the atmospheric phase cancelling out, thereby only leaving the sum of intrinsic phases of the object \citep{1958MNRAS.118..276J}. Closure phases are a probe for asymmetries, so deviations from values of 0\degr\ or 180\degr\ would indicate some deviation from centro-symmetry of the source. 

\subsection{$L$-band geometrical modelling (MATISSE)}\label{L-model}

\begin{figure}%[hbt!]
    \centering
    \includegraphics[trim=10 10 50 50, clip, width=\columnwidth]{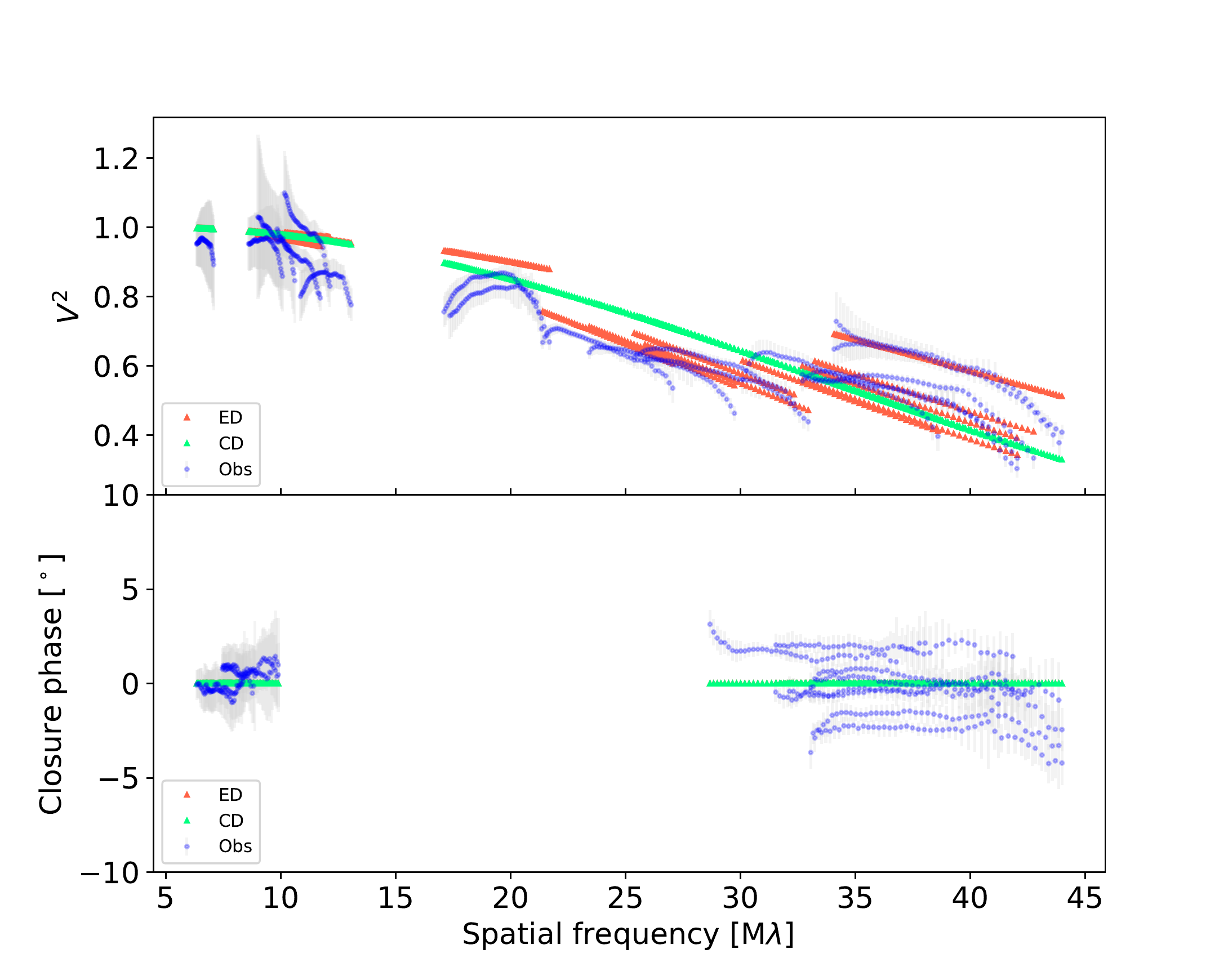}
    \caption{Observed (blue) and best fits for the $L$ band squared visibilities and closure phases. The models ED and CD refer to elliptical disk and circular disk, respectively.}
    \label{Fig-Lvis+phases}
\end{figure}

Previous geometrical modelling and preliminary imaging results in the $L$ band seem to suggest the presence of an elongated and tilted structure that enshrouds the central merger remnant \citep{2021A&A...655A.100M}. The previous VLTI observations in 2020 were obtained using only the large configuration. In the current study, V838 Mon was observed with the small (maximal baseline is $\approx30$ m) and large (minimal baseline $\approx$ 140 m) configurations in the $L$-band. A quick look at the squared visibilities as a function of spatial frequency suggests that the source is mostly resolved at the longer baselines (see Fig. \ref{Fig-Lvis+phases}). However, even with the most extended array configuration, V838 Mon is not completely resolved as the squared visibilities drop to a minimum of about 0.3. At the shortest baselines, V838 Mon is mostly unresolved, with the squared visibilities being at around 0.9. As a function of wavelength, though, it is clear that the squared visibilities do not show much variation, which suggests that no major chromatic effects are in play. At the very edge of the wavebands, the data was much noisier. We excluded it from our analysis and restricted ourselves to the wavelength range of 3--4 $\mu$m in the $L$ band. 

A qualitative look at the closure phases (see Fig. \ref{Fig-Lvis+phases}) indicates that they are 
mostly close to zero, with some very small deviations of about a few degrees (maximum $\sim$3\degr). This suggests a slight asymmetry in the system. Again, similar to the squared visibilities we do not see much variation in the closure phases as a function of wavelength, which suggests that the shape of the object is not changing over the covered wavelength range. 

To interpret the MATISSE data, we employed geometrical modelling, because the UV coverage of the MATISSE 2021/2022 data alone is insufficient for comprehensive imaging. We used the modelling software LITPRO provided by Jean-Marie Mariotti Center\footnote{\url{https://www.jmmc.fr/english/tools/data-analysis/oifits-explorer/}} (JMMC), which allows the user to fit simple geometrical models such as disks, point sources and ellipses to the observed visibilities and closure phases. LITPRO uses a $\chi^{2}$ minimization scheme to compute the model parameters and their error bars. For the purpose of fitting the V838 Mon data, we tried four models: a uniform circular disk (CD), an elliptical disk (ED), a circular Gaussian, and an elliptical Gaussian. With a reduced $\chi^{2}$ of 6, we find the elliptical models to better represent the data. The plots of the modelled and observed visibilities are shown in Fig.\,\ref{Fig-Lvis+phases}. Additionally, we are able to obtain precise estimates for the size and PA for the various models. These are listed in the Table \ref{model params}. The obtained parameters for size and orientation are in good agreement with what previous MATISSE observations had indicated \citep{2021A&A...655A.100M}. In \cite{2021A&A...655A.100M}, the elliptical disk model yielded a semi-major axis PA of $-40\degr \pm 6\degr$, an angular diameter of $3.28 \pm 0.18$ mas, and a stretch ratio of $1.40 \pm 0.1$. In this study, we get values of --41\fdg34$\pm$0\fdg49, $3.69 \pm 0.02$ mas, and $1.56 \pm 0.01$, respectively. It is evident that these parameters have remained largely the same between Jan/Mar 2020 and Oct 2021/Mar 2022. This means that the feature in the $L$ band is non-transient on a timescale of years. The absence of any change in PA also suggests that no dynamical variations have occurred in the post merger environment. Thus, it would seem that the disk-like feature is stable and long-lived in the $L$ band. This allowed us to combine the two datasets and attempt imaging (see  Sect.\,\ref{L_band_imaging}). 

%%%%%%%%%%%%%%%%%%%%%%%%%%%%%%%%%%%%%%%%%%%%%%%%
\subsection{$N$-band geometrical modelling (MATISSE)}\label{N-model}

\begin{figure}%[hbt!]
    \centering
    \includegraphics[trim=10 10 50 50, clip, width=\columnwidth]{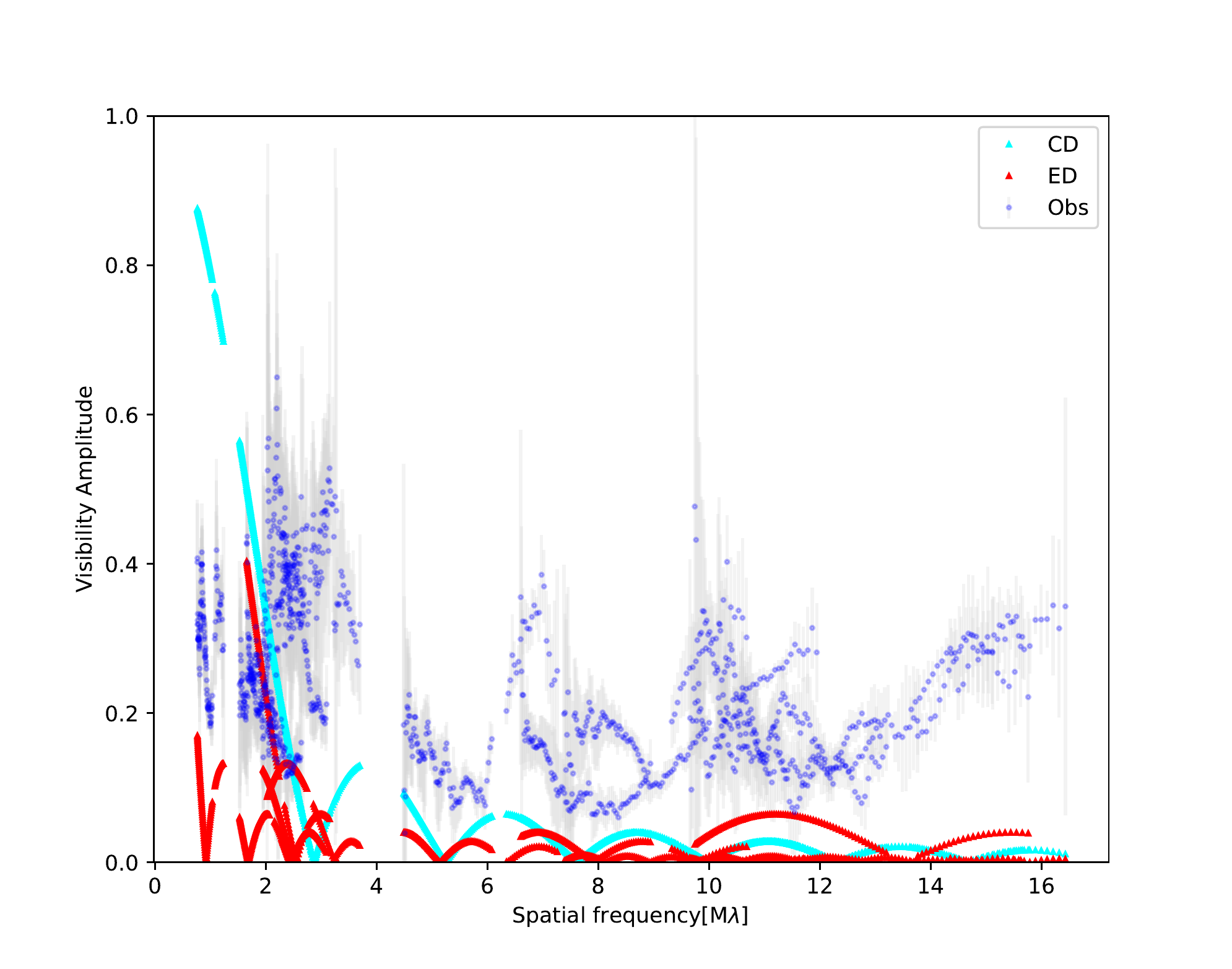}
    \caption{Observed (blue) and best fits for the $N$ band visibility amplitudes. The models ED and CD refer to elliptical disk and circular disk, respectively.}
    \label{Fig-Nvis}
\end{figure}

As stated previously, despite the generally poor quality of the $N$ band data due to thermal background and sensitivity limits, we were still able to obtain the correlated fluxes for V838 Mon for two of the three nights. Being measurements of the fringe amplitude, correlated fluxes are a proxy for fringe contrast and allow the determination of the source size. We modelled the $N$ band visibility amplitudes similar to the geometrical modelling we did for the squared visibilities in other bands. Given the scarce number of data points, we used simple geometries to model the observables. In the $N$ band, our goal was to obtain a size measurement for the structure and also constraints on its ellipticity. We used two models, a uniform disk and an elliptical disk. The uniform disk model fit yields an angular diameter of $87.73 \pm 0.01$ mas, while the elliptical disk model yields an angular diameter of $48.70 \pm 2.40$ mas, a PA of 5\fdg03$\pm$0\fdg70, and a stretch ratio of $5.68 \pm 0.31$. These fits are shown in Fig.\,\ref{Fig-Nvis}. If we consider the size estimate obtained from the elliptical disk model (semi-major axis of $\sim 300$ mas), then it would seem that the most extended structure surrounding V838 Mon exists in the $N$ band. The PA points to a north-south elongation in the structure, which is drastically different from that seen in other bands. This means that the $N$ band structure is a distinct feature that may not be dynamically related to structures seen in the $HKL$ bands.

%%%%%%%%%%%%%%%%%%%%%%%%%%%%%%%%%%%%%%%%%%%%%%%%
\subsection{$M$-band geometrical modelling (MATISSE)}\label{M-model}

\begin{figure}%[hbt!]
    \centering
    \includegraphics[clip, width=\columnwidth]{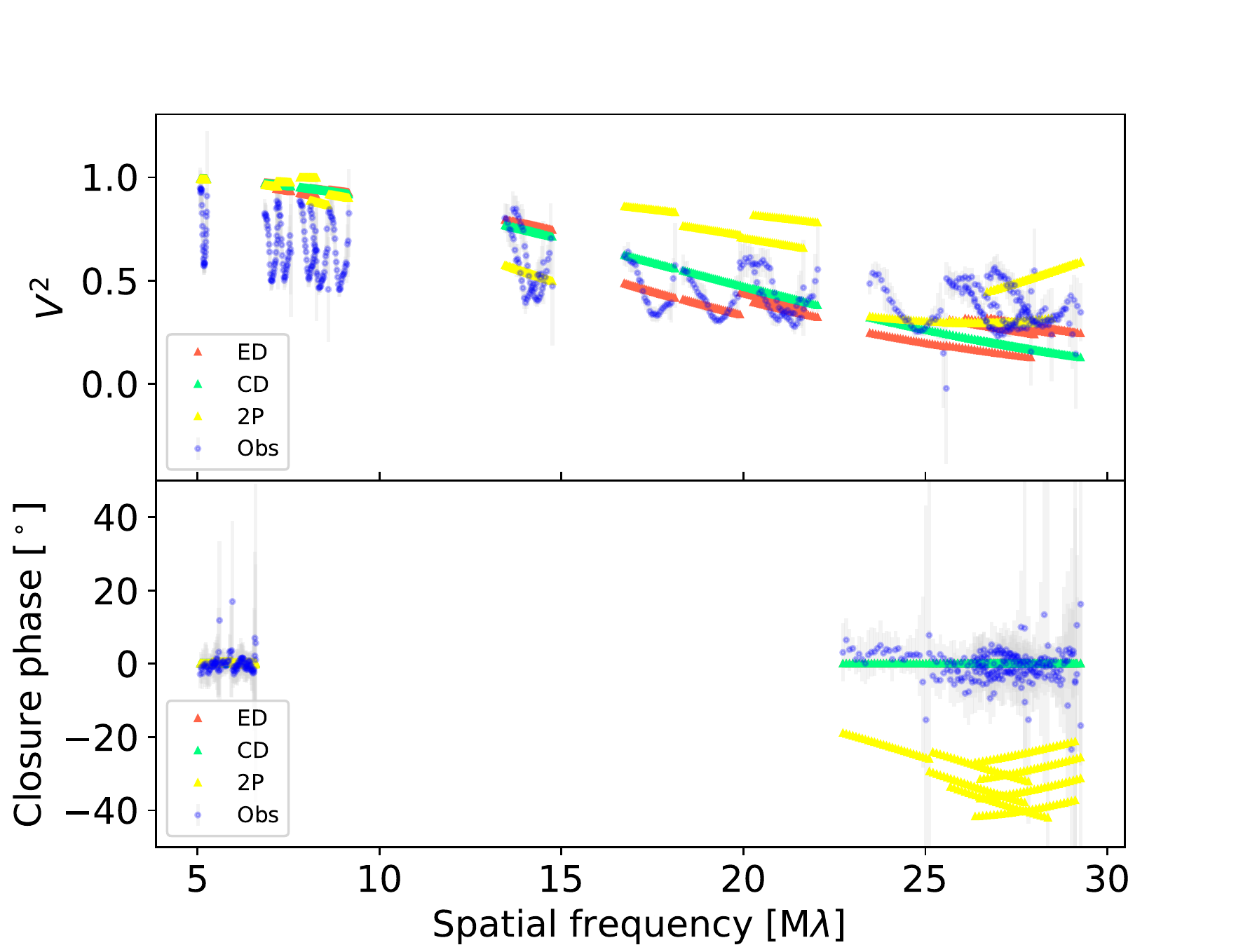}
    \caption{Observations (blue) and best fit models for the $M$ band squared visibilities and closure phases. Models ED, CD and 2P refer to elliptical disk, circular disk, and two point sources, respectively.}
    \label{Fig-Mvis+phases}
\end{figure}

For the first time, we were able to observe V838 Mon in the $M$ band using MATISSE. However, due to the scarce data in this band, we did not attempt any image reconstruction. As can be seen in Fig. \ref{Fig-Mvis+phases}, the shape of the squared visibilities as a function of spatial frequency shows a sinusoidal-like modulation. It is prominent across all spatial frequencies, but becomes more noticeable at longer baselines. The amplitude variation lies in the range 0.3--0.4, and the frequency of these modulations increases at lower spatial frequencies. The $M$-band closure phases are mostly close to zero. However, at larger spatial frequencies they tend to deviate by a few degrees (maximum $\sim 5 \degr$), thereby hinting towards asymmetries similar to what we see in the $L$-band. 

\begin{figure}%[hbt!]
    \centering
    \includegraphics[clip, width=\columnwidth]{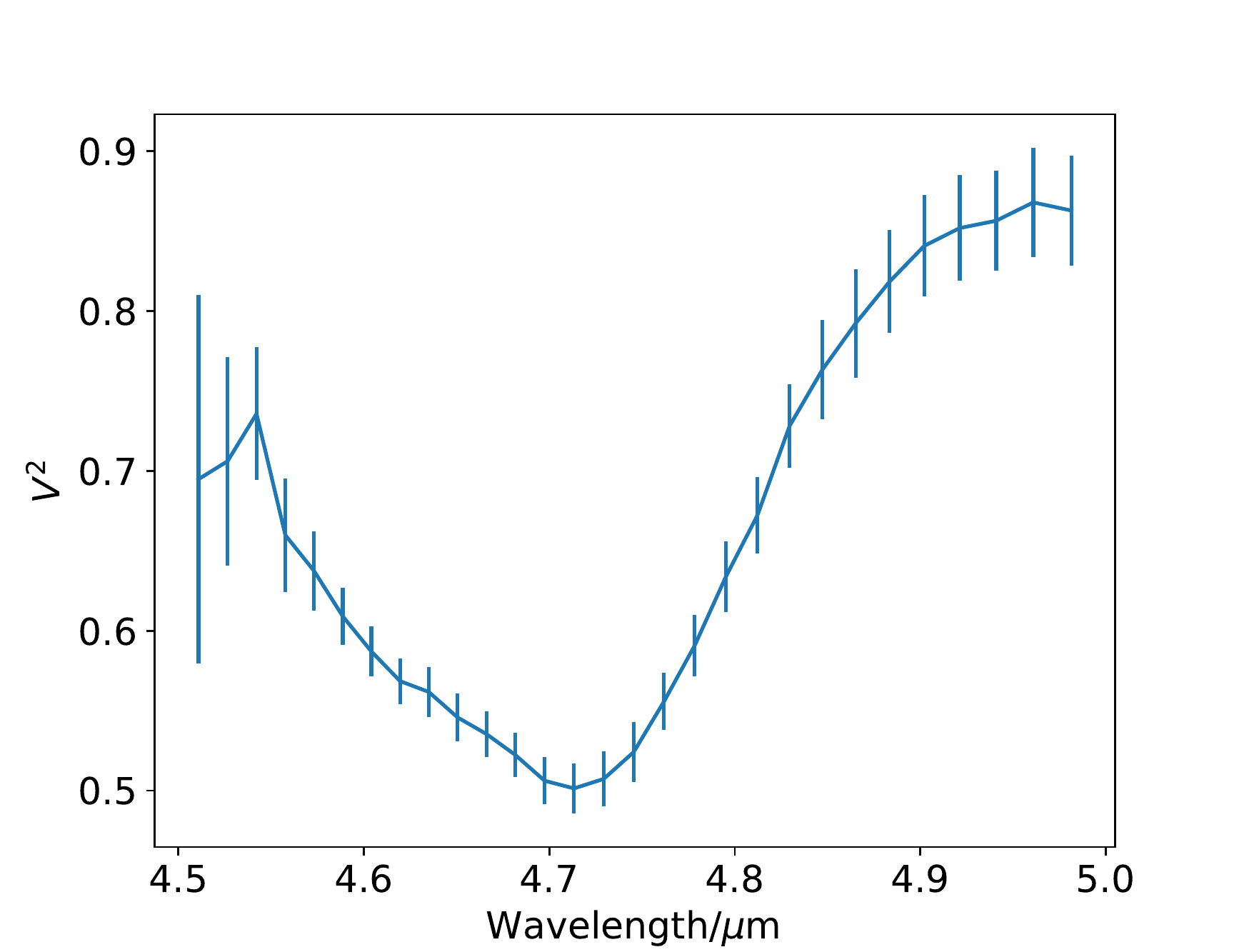}
    \caption{Observed squared visibilities as a function of wavelength near the 4.7 $\mu$m dip. Data for all baselines were averaged.}
    \label{Fig-visVswav}
\end{figure}

\begin{figure}%[hbt!]
    \centering
    \includegraphics[clip, width=\columnwidth]{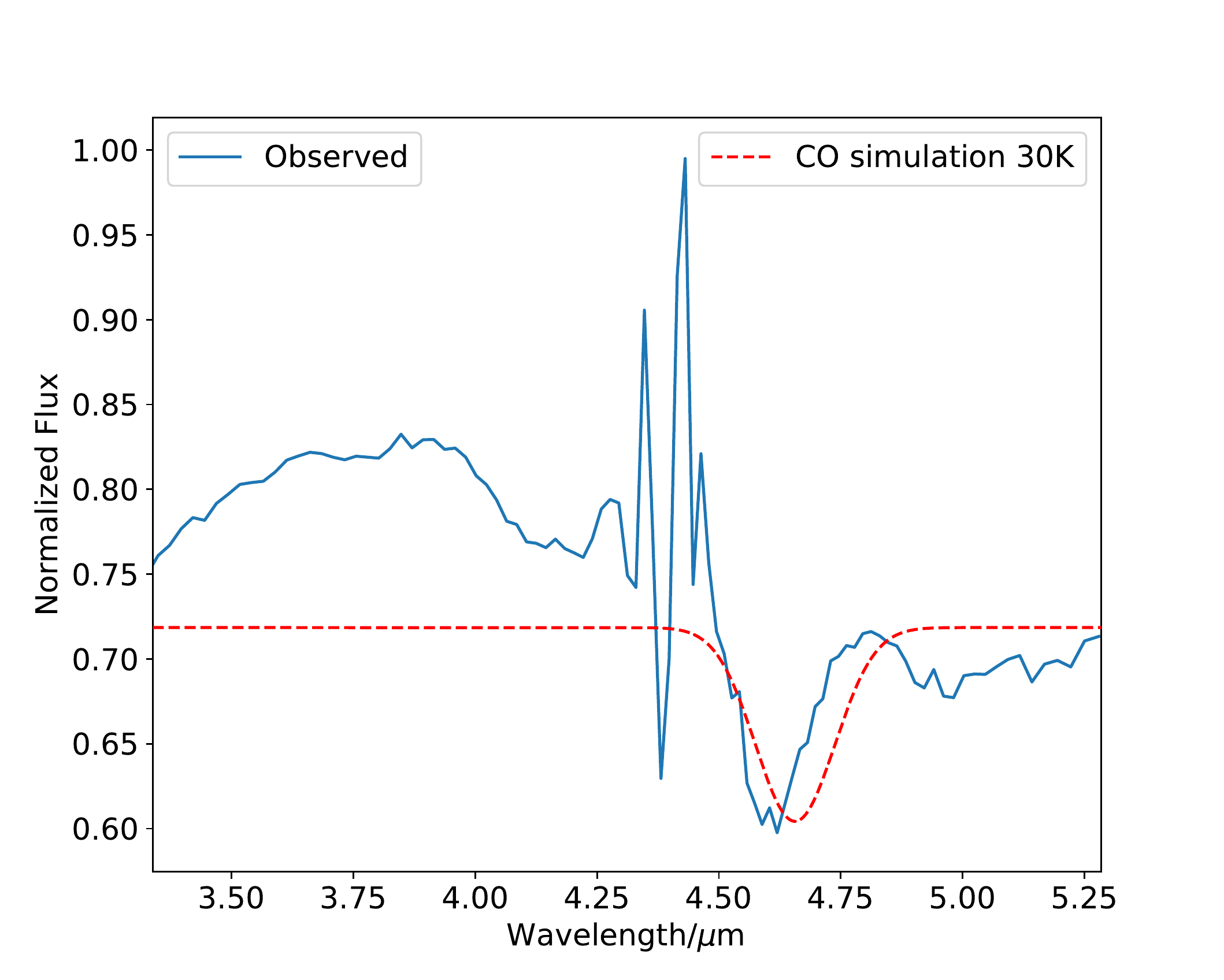}
    \caption{Observed $LM$-band spectrum (solid line) is compared to a simple simulation (dashed line) of absorption at the CO fundamental band near 4.6 $\mu$m and at a gas temperature of 30 K. The simulated spectrum was smoothed to the spectral resolution of MATISSE.}
    \label{M band CO}
\end{figure}

The visibilities vary also as a function of wavelength and show a broad depression near 4.7\,$\mu$m. This is illustrated in Fig.\,\ref{Fig-visVswav} for visibilities averaged over all baselines. A minimum is reached at 4.7\,$\mu$m which is close to the fundamental band of CO (Fig.\,\ref{M band CO}). The presence of the feature explains the sinusoidal variations seen in Fig. \ref{Fig-Mvis+phases}. Closure phases do not seem to show much wavelength dependence throughout this particular waveband. Our flux calibrated spectra, however, readily show an absorption feature at around 4.6 $\mu$m, which we identify as the fundamental band of CO. In Fig.\,\ref{M band CO}, we compare the entire calibrated $LM$ spectrum to a simulation of CO $\Delta v = \pm 1$ absorption from an optically thin slab at thermal equilibrium and at a gas temperature of 30 K. The match in wavelength is not perfect, but can be explained by the complex pseudo-continuum baseline on both sides of the feature. The observed feature may not be pure absorption. Also, telluric absorption of water makes this part of the band particularly difficult to observe and calibrate. %Nevertheless, the spectrum of V838 Mon in $L$ band is similar to spectra of some cool giants with dense envelopes, for instance to TX Cam (a MIRA of spectral type M8--M10) in The Cassini Atlas Of Stellar Spectra\footnote{\url{http://www.physics.usyd.edu.au/sifa/caoss/}} \citep{cassini}. 
Our identification of the CO band is strengthened by observations of the same band in V838 Mon 2005 at a much higher spectral resolution \citep{geballe}, and its presence in other red novae, for instance in V4332 Sgr \citep{2004ApJ...615L..53B}.

Although the $L$-band spectrum shows clear signatures of circumstellar molecular features, limited by the small number of measured visibilities, we are currently able to make basic size measurement only by combining data in the entire observed band.  We used a wide number of achromatic models, which include a circular disk (CD), circular Gaussian (CG), elliptical disk (ED), elliptical Gaussian (EG) and a two point source model (2P). The fits to these models and their details are given in Table \ref{model params}, while the models of the visibilities and closure phases are also shown in Fig.\,\ref{Fig-Mvis+phases}.  None of the models are able to reproduce exactly the modulation shown by the visibilities. Disk-like models with a size of 4--6\,mas fit the data much better than a double source. The derived size most likely represents the dominant circumstellar component.

\subsection{$K$-band geometrical modelling (GRAVITY and MYSTIC)}\label{K-model}

%%%%%%%%%%%%%%%%%%%%%%%%%%%%%%%%%%%
 
\begin{figure}%[hbt!]
    \centering
    \includegraphics[trim=10 0 30 0, clip, width=\columnwidth]{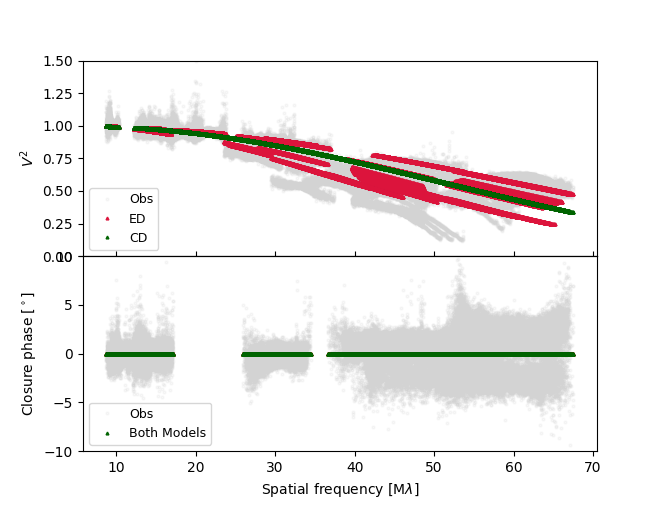}
    \caption{$K$ band squared visibilities and closure phases measured with GRAVITY (gray). Best-fit geometrical models for elliptical disk (ED) and circular disk (CD), are also shown (both have zero closure phases).}
    \label{Fig-Kvis+phases}
\end{figure}

\begin{figure}%[hbt!]
    \centering
    \includegraphics[clip, width=\columnwidth]{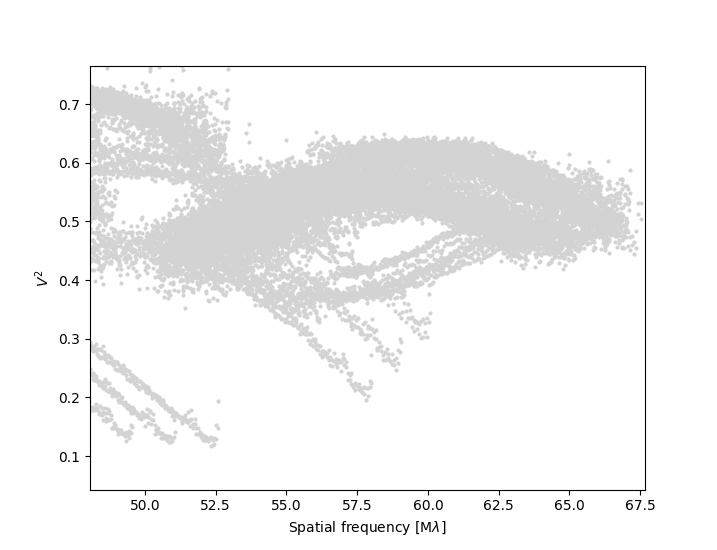}
    \caption{Zoomed-in section on some of the `bump' features in the squared visibilities from Fig.\,\ref{Fig-Kvis+phases}.}
    \label{Fig-zoom}
\end{figure}

\begin{figure}%[hbt!]
    \centering
    \includegraphics[clip, width=\columnwidth]{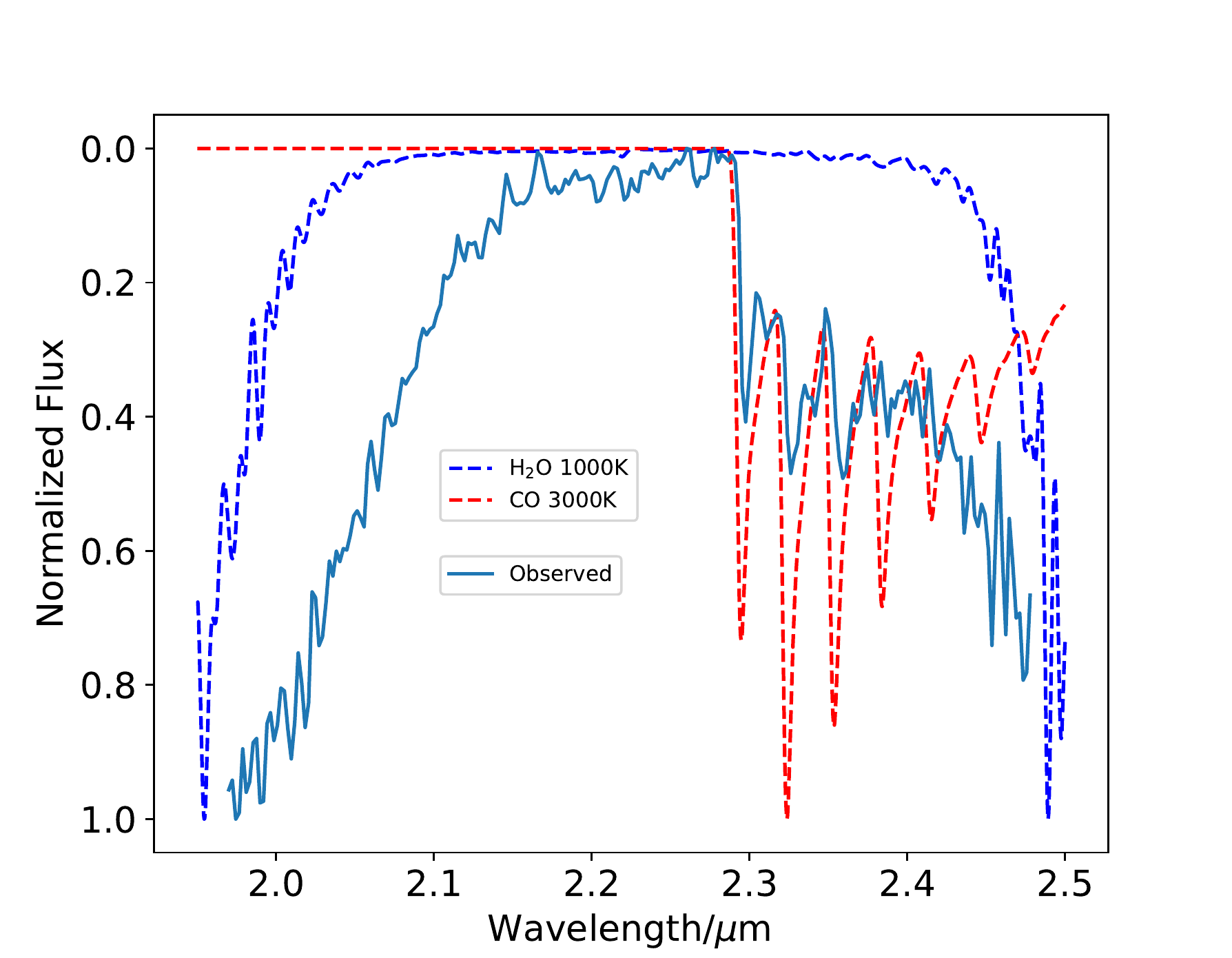}
    \caption{Synthetic spectra generated for H$_{2}$O at 1000\,K and for CO at 3000\,K are compared to the observed $K$ band spectrum from GRAVITY over the entire observed band. The synthetic spectra are arbitrary scaled. The simulation does not include high-opacity effects, which may result in the poor fit to the observed saturated water band near 2 $\mu$m. The synthetic spectra were convolved to the spectral resolution of the GRAVITY resolution.}
    \label{Fig-comb-spec}
\end{figure}

\begin{figure}%[hbt!]
    \centering
    \includegraphics[clip, width=\columnwidth]{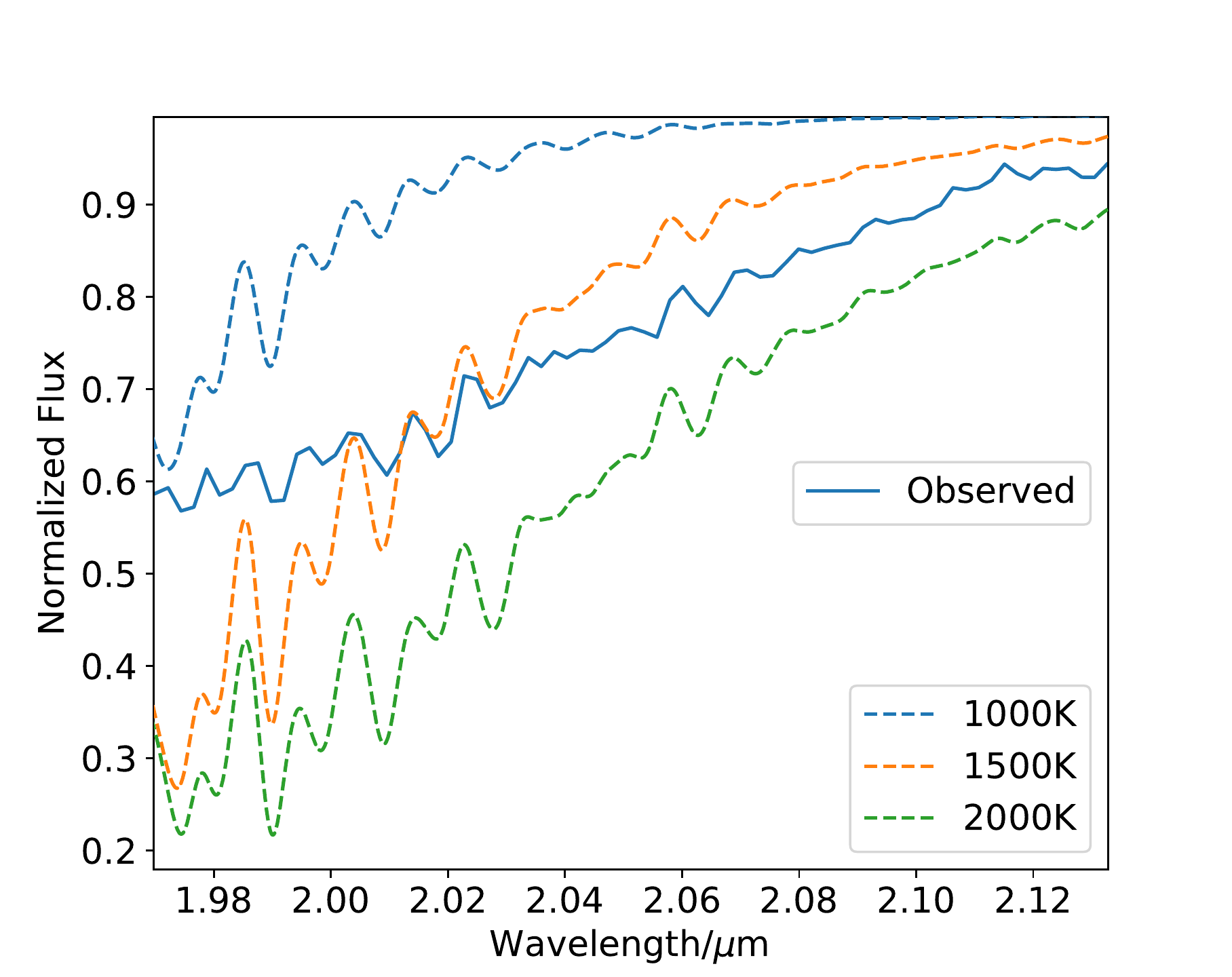}
    \caption{Synthetic spectra of water generated at various temperatures are compared to the observed spectrum (blue).}
    \label{Fig-water-spec}
\end{figure}

 The GRAVITY $K$ band observations for V838 Mon are the most extensive of all the VLTI observations that we have obtained so far. Observations on all three array configurations, including the intermediate configurations, were successfully executed, which resulted in good sampling of the UV plane and allowed us to perform image reconstruction (see Sec. \ref{K_band_imaging}). We obtained 15 out of the requested 18 observing runs. Results are shown in Fig. \ref{Fig-Kvis+phases}. The source is well resolved at the longest baselines, as the squared visibilities fall to a minimum of 0.1. The longer baselines also reveal smaller peak-like features, which point to a complex morphology in the $K$ band that cannot be explained by just a simple disk or a Gaussian model (shown more clearly in Fig.\,\ref{Fig-zoom}). The squared visibilities do not change considerably with wavelength. This suggests that the shape of the structure is not wavelength dependent throughout the entire $K$ band. We do see, however, subtle spectral effects near 2 and 2.3 $\mu$m which we assign to water and CO absorption (see Figs. \ref{Fig-comb-spec} and \ref{Fig-water-spec}). The closure phases seem to scatter around zero by a few degrees and reach a maximum of $\sim$5\degr). The root-mean-square (RMS) scatter is 2\degr\ which suggests that the observed deviations in the closure phases can be attributed mostly to observational noise (within 4$\sigma$) and maybe some very small asymmetries present in V838 Mon. They also do not seem to vary with wavelength, thus indicating a non-variable morphology in the $K$ band. 
 
 Although the $K$ band data were used extensively for imaging, we nevertheless thought it prudent to also model the visibilities directly using simple geometrical models in order to obtain constraints on the size of the structure. The best fit parameters for all models are shown in Table \ref{model params}. The size of the source lies in the range 1.2--2.6 mas, depending on the choice of model. The value of the PA in the $K$ band is similar to measurements in the other bands.
 
\begin{figure}%[hbt!]
    \centering
    \includegraphics[clip, width=\columnwidth]{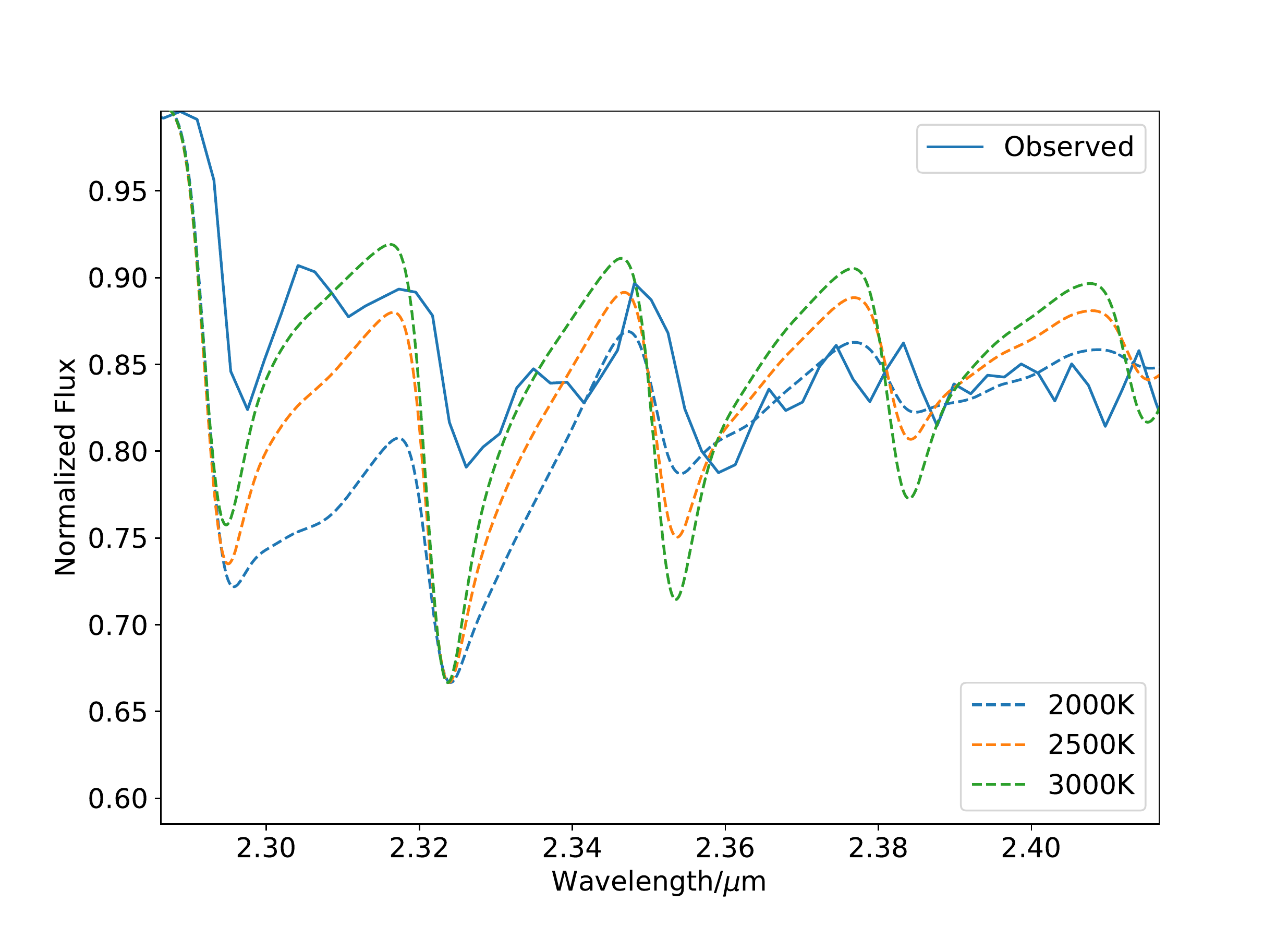}
    \caption{Synthetic spectra generated for the first-overtone CO bands at various temperatures. The solid blue line is the observed spectrum.}
    \label{Fig-CO-spec}
\end{figure}

We also were able to obtain $K$ band spectra with a medium resolution (R$\sim$500). The spectrum was calibrated using the star HD\,54990. This resulted in the removal of the major telluric feature caused by H$_{2}$O near 2 $\mu$m, leaving us with absorption bands in the range 2.3 -- 2.35 $\mu$m and less intense absorption features near 2 $\mu$m. We identified these as CO and H$_{2}$O features, respectively. To ensure that this identification was accurate, we modelled the absorption at various temperatures using {\tt pgopher} \citep{pgopher} and spectroscopic data from the EXOMOL database \citep{exomol}. The models represent a simple slab of gas in local thermodynamic equilibrium (i.e., characterized by a single temperature) and optically thin absorption. Observed spectra are compared to the simple simulations in Figs.\, \ref{Fig-comb-spec}, \ref{Fig-water-spec}, and \ref{Fig-CO-spec}. After experimenting with a wide variety of temperatures for both species, we find that a model with a CO temperature of 3000\,K and an H$_{2}$O temperature of 1000\,K best represent the data, but due to the poor spectral resolution these are only very rough constraints. The absorption may arise in the photosphere or very close to it in the circumstellar gas \citep[cf.][]{geballe}. We use geometrical modelling to determine the sizes of the source when the visibilities are limited to spectral regions where molecular absorption is strongest. For CO, we restrict the GRAVITY data to 2.29--2.37 $\mu$m, while for water 1.97--2.07 $\mu$m was selected. By fitting circular disk models to each of these datasets, we obtain angular diameters of $2.1 \pm 0.1$ mas and $2.0 \pm 0.1$ mas, respectively. With the uncertainties, these are consistent with the size obtained for the entire band.  

\begin{figure}%[hbt!]
    \centering
    \includegraphics[trim=20 0 50 0, clip, width=\columnwidth]{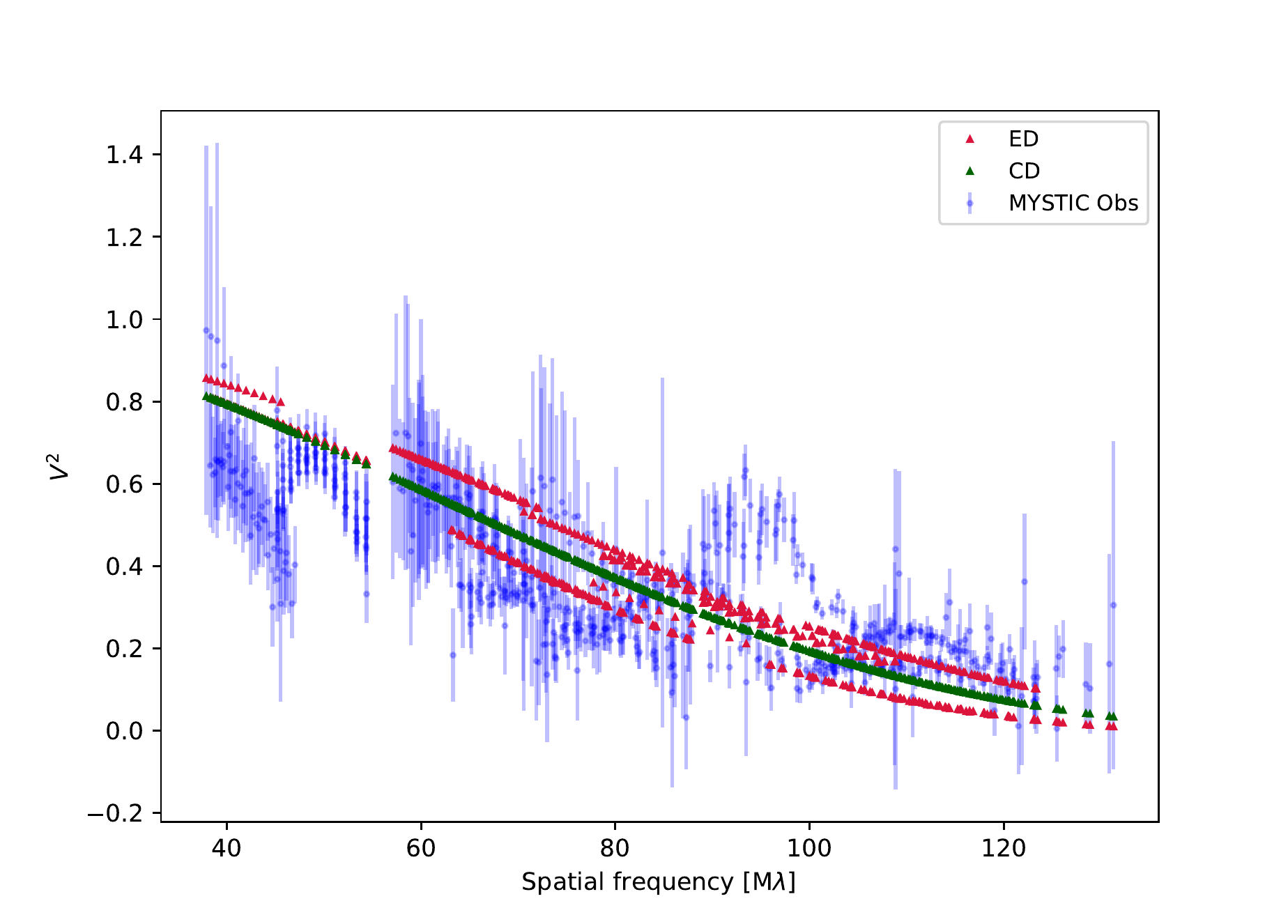}
    \caption{$K$ band squared visibilities as measured with MYSTIC (blue). Best-fit geometrical models for elliptical disk (ED) and circular disk (CD) are also shown.}
    \label{Fig-mystic-vis}
\end{figure}

We also obtained a single night of MYSTIC $K$ band observations and geometrically modelled the visibilities with a uniform disk and an elliptical disk model. The results are shown in Fig.\,\ref{Fig-mystic-vis}. The uniform disk model yields an angular diameter of $1.57 \pm 0.01$, mas while the elliptical disk model yields an angular diameter of $1.72 \pm 0.01$\,mas, a PA of $-41.6 \pm 1.5$\degr, and a stretch ratio of $1.28 \pm 0.01$. Due to the longer CHARA baselines, V838 Mon is much better resolved, which yields a smaller angular diameter than that obtained with just the GRAVITY observations. Additionally, we also combined both the GRAVITY and CHARA squared visibilities and performed a combined uniform disk fit to the dataset, which gave an angular diameter estimate of $1.94 \pm 0.01$\,mas. This value is essentially the same as the one obtained from fitting only the GRAVITY squared visibilities, which suggests that combining them with the MYSTIC squared visibilities does not result in a significant change in the size estimate for V838 Mon. 

The mean of the MYSTIC closure phases is $\sim$2\degr\ with an RMS of $\sim$11\degr, which strongly hints towards the presence of asymmetries. Unlike the GRAVITY dataset, the MYSTIC observations lasted only a single night because of which we could only geometrically model the closure phases. The results of this modelling are presented in Sect.\,\ref{MYSTIC_CPS}.

%\subsection{Size of the CO and H$_2$O regions}
%The $K$ band spectra of V838 Mon show that the most prominent spectral features are the CO absorption lines followed by the much weaker spectral features near $2 \mu$m that we identify as water absorption lines. Past studies have also shown the presence of water and CO in the post-merger environment. 

%\begin{figure}%[hbt!]
 %   \centering
  %  \includegraphics[clip, width=\columnwidth]{H2O30CO3000.pdf}
  %  \caption{Synthetic spectra generated for H$_{2}$O at 30 K and CO at 3000 K.}
   % \label{Fig-H2O30CO3000-spec}
%\end{figure}

%\begin{figure}%[hbt!]
 %   \centering
  %  \includegraphics[clip, width=\columnwidth]{H2O3000CO300.pdf}
   % \caption{Synthetic spectra generated for H$_{2}$O at 3000 K and CO at 300 K.}
   % \label{Fig-H2O3000CO30-spec}
%\end{figure}

%\begin{figure}%[hbt!]
 %   \centering
  %  \includegraphics[clip, width=\columnwidth]{H2O1000CO30.pdf}
  %  \caption{Synthetic spectra generated for H$_{2}$O at 1000 K and CO at 30 K.}
  %  \label{Fig-H2O1000CO30-spec}
%\end{figure}

% We obtained additional $K$ band data with the MYSTIC instrument at CHARA. We modelled the latter with a disk model in order to see how the size estimation in the $K$ band changes by using the longer baselines of CHARA. The resulting diameter, assuming a circular disk model, is $1.57 \pm 0.01$ mas, that is, slightly smaller than the one estimated with GRAVITY data alone ($1.94 \pm 0.01$ mas). This is because the longest CHARA baselines better resolve V838 Mon in the $K$ band compared to the VLTI.   

\subsection{$H$-band geometrical modelling}\label{sec-Hgeo}
\begin{figure}%[hbt!]
    \centering
    \includegraphics[trim=10 0 40 20, clip, width=\columnwidth]{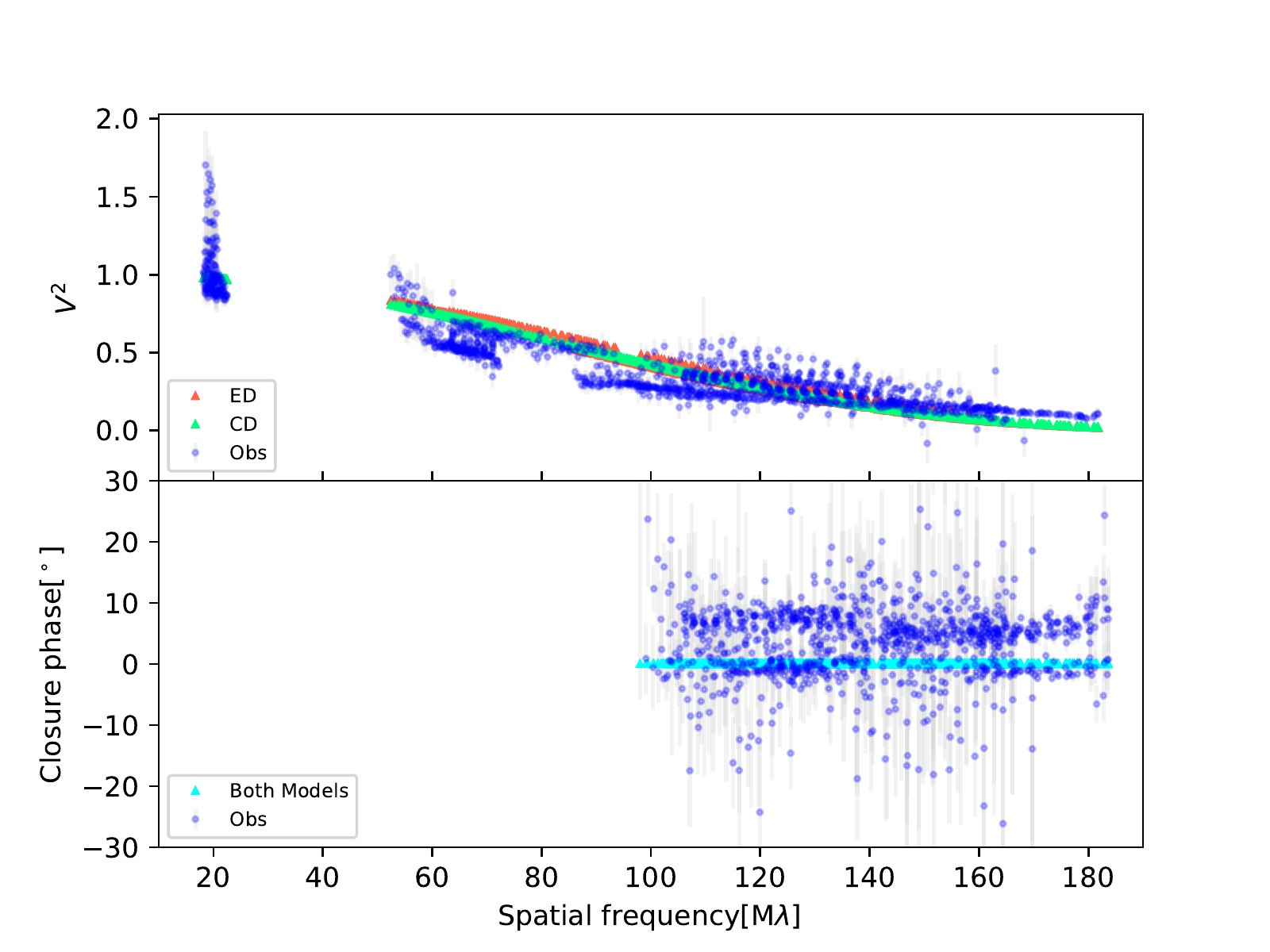}
    \caption{Observations (dark blue) and best fit models for the $H$ band squared visibilities and closure phases. Circular (CD) and elliptical (ED) disk models are shown.}
    \label{fig-H-band}
\end{figure}

CHARA MIRC-X observations allowed us to observe V838 Mon in the $H$ band as well. Previous measurements in the $H$ band were taken in 2012 using the AMBER instrument, and prior to that in 2004 \citep{2005ApJ...622L.137L} using the Palomar Testbed Interferometer (PTI). These recent CHARA measurements serve as a direct follow-up to the previous $H$ band observations. The squared visibilities, shown in Fig.\,\ref{fig-H-band}, suggest that the source is resolved at the longest CHARA baselines ($\lesssim$331 m), as the visibilities fall to a minimum of 0.1. The closure phases split into two groups. One is scattered around null phase, and the other, more populous group, is centered at phases $\sim$6\degr, thereby hinting towards deviations from centro-symmetry in some parts of the remnant. This pattern is observed at all spatial frequencies and requires a more complex source structure, which we discuss in Sect. \ref{H_band_imaging}.

%We fit the visibilities and closure phases in LITPRO in order to roughly constrain the size and shape of the source in the $H$ band. 
%Measuring the size of V838 Mon in the $H$ band is particularly important, since comparing the current value to previous estimates in the literature can help us understand the dynamic evolution of the merger product. 
%We fit two models to the observables in LITPRO, a uniform disk and an elliptical disk (see Table \ref{model params}), which yield a size of 1.2 mas. They, however, fail to explain the non-zero closure phases. This suggests that at least in the $H$ band V838 Mon does possess significant asymmetries within its circumstellar structure given the relatively grater closure phase deviations (maximum deviation $\sim 20\degr$). Such deviations are most likely to be caused by the source morphology itself rather than random scatter. Thus any detailed probing of these asymmetries requires high resolution image reconstruction which we also performed (see Section \ref{H_band_imaging}). The values of the fitted parameters for both the models can be found in Table \ref{model params}, for example for a circular disk model the angular diameter turns out to be $1.180 \pm 0.004$ mas.

%ADD  SUMMARY SKETCH WITH ALL ELLIPTICAL FITS SHOWN TO SCALE.

\section{Image reconstruction} \label{imaging}

\subsection{$L$ band image reconstruction}
\label{L_band_imaging}
For the purpose of image reconstruction, we combined the 2021/2022 MATISSE $L$ band data set with the one from 2020 used by \cite{2021A&A...655A.100M}, since no significant variability occurred in the source over the course of the year. This is further confirmed by our modelling of the recent visibilities and phases presented in Sect.\,\ref{L-model}. Since both of the datasets, except for one, consist of observations made with the large configuration, any reconstructed image will be sensitive to compact features in V838\,Mon. The image was constructed using the image reconstruction algorithm Multi-aperture Image Reconstruction Algorithm MIRA \citep{2008SPIE.7013E..1IT} which uses a likelihood cost function minimization method. For the $L$ band image, we set a pixel size of 0.1 mas, which is smaller than the theoretical best resolution ($\lambda_L/2B_{\rm max}\approx 2$\,mas). This was done as an attempt to reveal any finer substructure in V838 Mon at this band. A FoV of 5\,mas was chosen, taking into consideration the size of the $L$ band structure inferred from geometrical modelling (see Table \ref{model params}). The regularization scheme we employed was the hyperbolic regularization with a regularization parameter of $10^{6}$. These regularization settings resulted in images with the lowest $\chi^{2}$ values.       

\begin{figure}%[hbt!]
    \centering
    \includegraphics[trim=5 0 35 30, clip, width=\columnwidth]{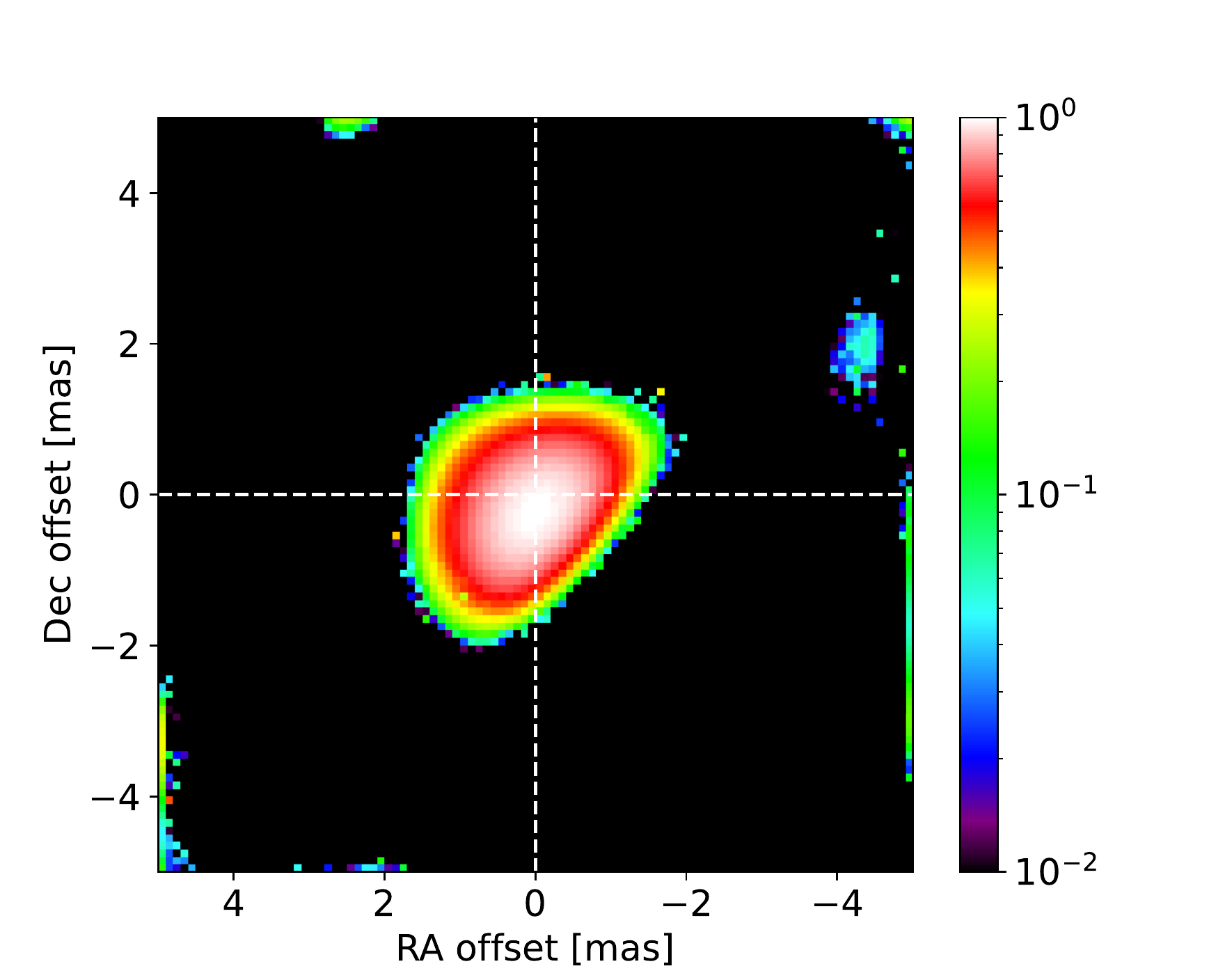}
    \caption{Image reconstruction of V838 Mon done in the $L$ band using the combined 2020--2022 MATISSE data. Flux is in a logarithmic scale in arbitrary units.}
    \label{MATISSE-combined}
\end{figure}

The resulting image displayed in Fig.\,\ref{MATISSE-combined} shows an asymmetrical, elliptical feature. The geometric orientation is along the PA that we obtained from our modelling presented in the prior sections.

\subsection{$K$ band image reconstructions}

\label{K_band_imaging}

\begin{figure}%[hbt!]
    \centering
    \includegraphics[clip, width=\columnwidth]{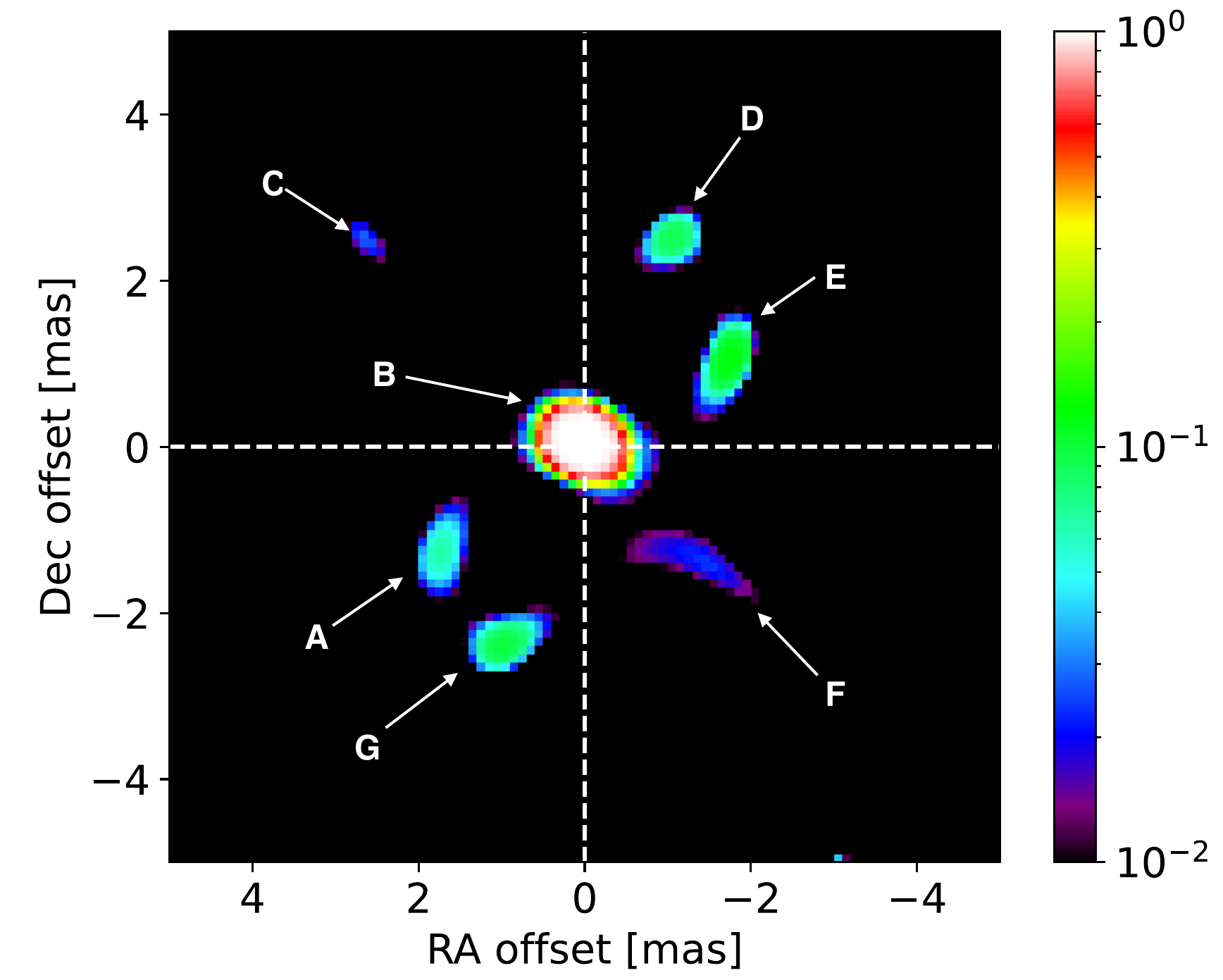}
    \caption{Reconstructed $K$ band image of V838 Mon obtained using the MIRA imaging algorithm. Flux level is logarithmic, and the units are arbitrary.}
    \label{MIRA_GRAVITY}
\end{figure}

\begin{figure}%[hbt!]
    \centering
    \includegraphics[trim=10 0 0 30, clip, width=1.1\columnwidth]{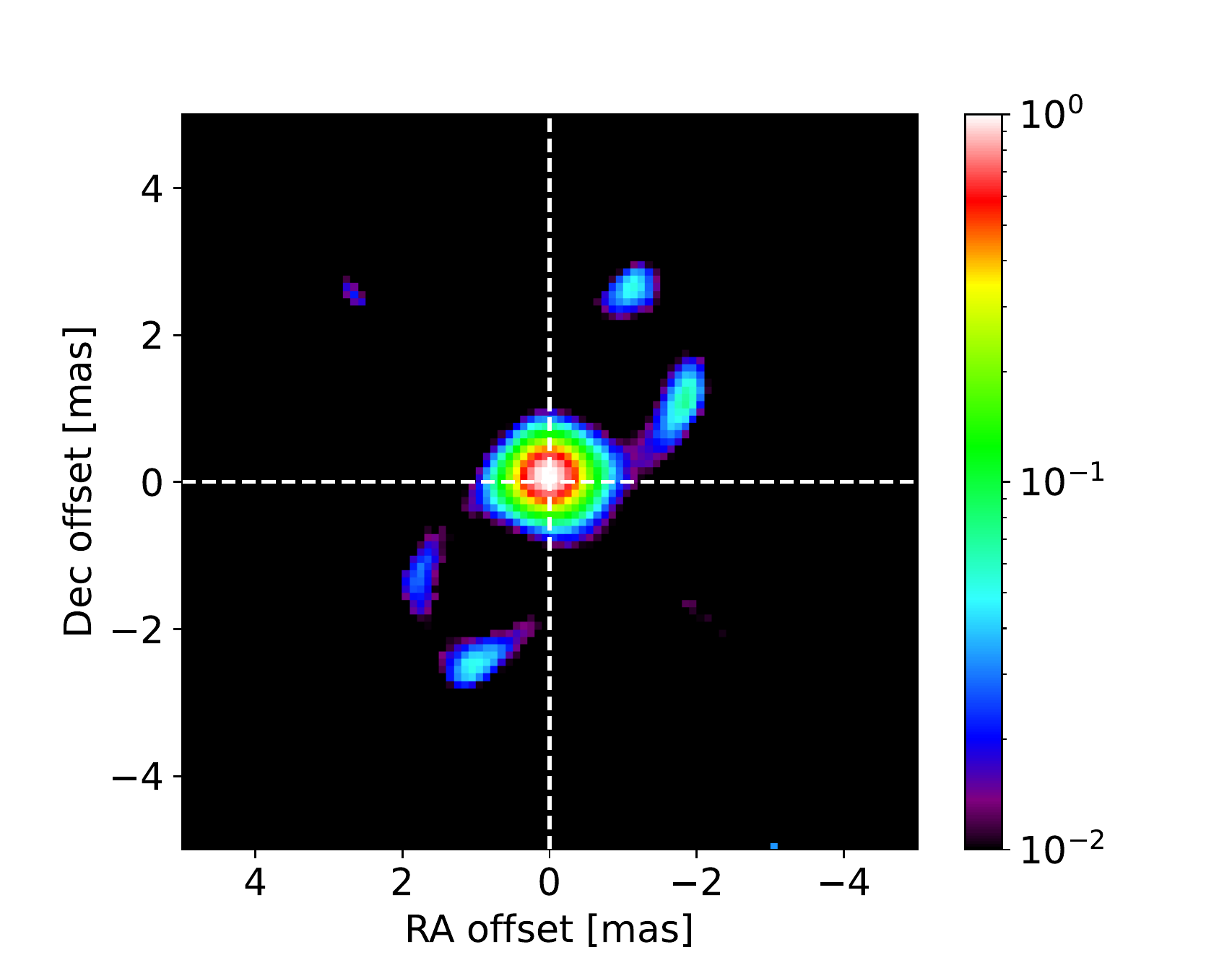}
    \caption{Reconstructed $K$ band image of V838 Mon obtained using the SQUEEZE imaging algorithm. }
    \label{SQUEEZE_GRAVITY }
\end{figure}

\begin{figure}%[hbt!]
    \centering
    \includegraphics[trim=10 0 0 30, clip, width=1.1\columnwidth]{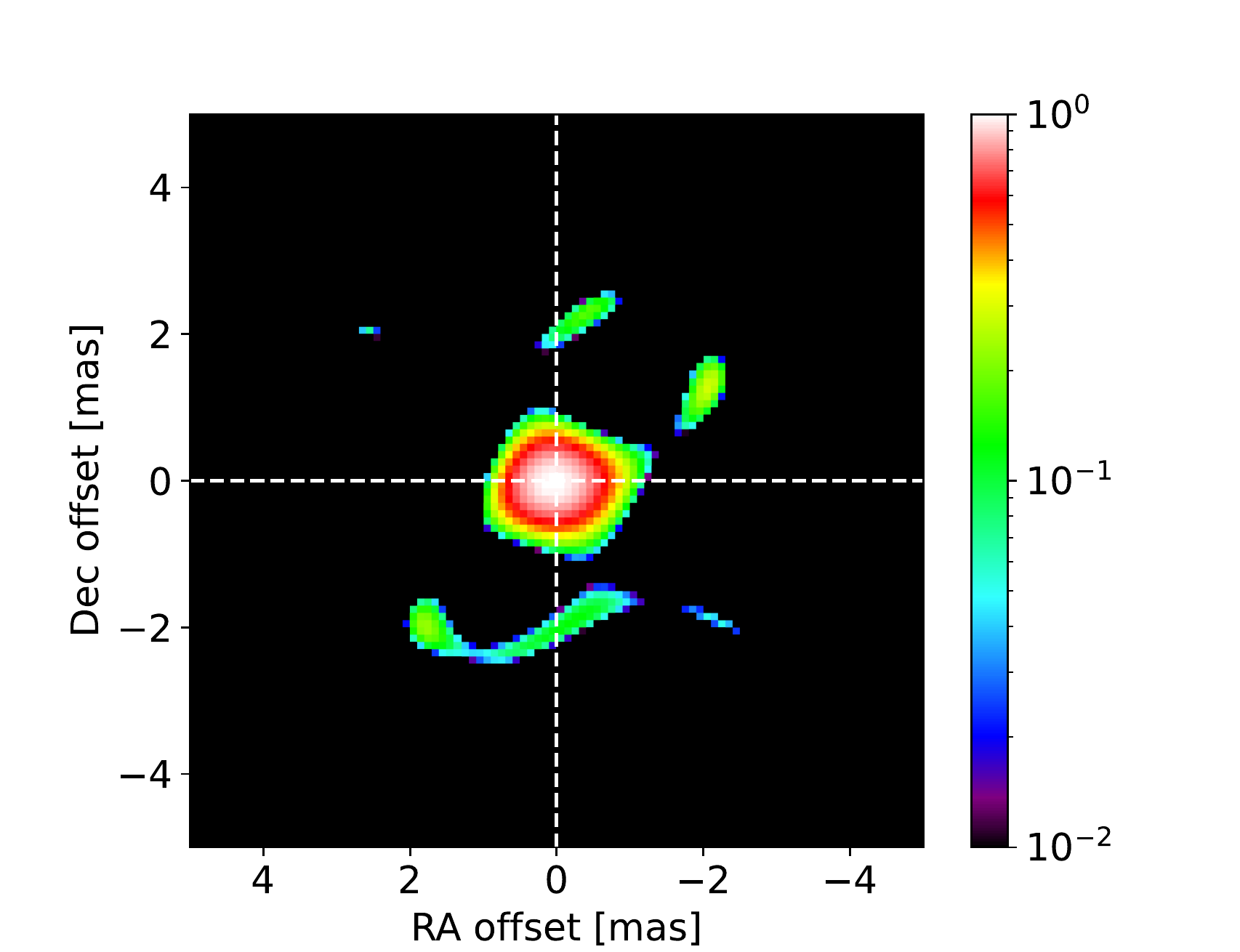}
    \caption{Reconstructed $K$ band image of V838 Mon obtained using the chain imaging method with the MIRA imaging algorithm. }
    \label{chain_image}
\end{figure}

Given the comprehensive nature of our GRAVITY observations, we were able to perform detailed image reconstruction of V838\,Mon from the measured interferometric observables in the $K$ band. For this purpose, we used two separate image reconstruction algorithms, SQUEEZE \citep{2010SPIE.7734E..2IB} and MIRA \citep{2008SPIE.7013E..1IT}. Both these algorithms attempt to solve for the best image given the data and a model of the image. They employ different methodologies and minimization techniques from one another. Thus, by using multiple algorithms to reconstruct an image and comparing them to one another, one can distinguish between authentic features present in a source and artifacts that could have resulted from the different algorithms. For the purpose of our image reconstruction, we limited our FoV to 10 mas and set the pixel size to 0.1 mas, thus making the image 100 by 100 pixels. This choice for FoV and pixel size was adopted in order to super resolve any potential sub-mas features. With MIRA, just like in the $L$ band, we used the hyperbolic regularization and found that the $\chi^{2}$ was minimized with a regularization parameter of $10^6$. While with the SQUEEZE algorithm, we used  total variation and found the $\chi^{2}$ to be minimized with a parameter value of $10^3$. Total variation with a high enough regularization parameter value is identical to the hyperbolic regularization scheme. The values for the regularization parameters were determined by reconstructing images for a wide range of parameter values and by noting the value at which no further reduction in $\chi^{2}$ takes place. This parameter is significant since it sets the weight for the image model against the data when the minimization scheme is implemented. We combined all the GRAVITY OIFITS files and merged them into a single file by using OI-Tools\footnote{https://github.com/fabienbaron/OITOOLS.jl/tree/main/demos}. This yielded a single OIFITS file that was then used as the input file in the image algorithms. The number of iterations was set at 500. Finally, a randomly generated image was used as the starting image for all of our reconstruction attempts. The reconstruction results with both MIRA and SQUEEZE are shown in Fig.\,\ref{MIRA_GRAVITY} and \ref{SQUEEZE_GRAVITY } respectively. We performed image reconstruction using a different strategy as well, whereby the combined GRAVITY data OIFITS file was binned into smaller wavelength increments of 0.05\, $\mu$m. The image in the first wavelength bin in this process was constructed with a random starting image. However, in subsequent reconstructions for the other bins, the starting image used would be the image obtained for the preceding bin. Using this chain imaging method helped us to significantly decrease the value of the reduced $\chi^{2}$, i.e. from $\approx 150$ to $\approx 8$ in the final image. Thus, the final image is at the endpoint of the waveband. The result of this strategy is displayed in Fig.\,\ref{chain_image}.

The resulting images (Figs.\,\ref{MIRA_GRAVITY} and \ref{SQUEEZE_GRAVITY }) from both the algorithms show a similar morphology in the $K$ band. At 0.1 mas, multiple features are immediately noticeable. These include the clump-like features (labelled as A, B, D, E, and G in Fig.\,\ref{MIRA_GRAVITY}) that are roughly distributed along an axis at the PA value ($\sim$--40\degr) which we have obtained from geometrical modelling and which is similar to the PA found for the disk-like feature found in V838 Mon in other bands. Furthermore, we also see two narrow linear features (labelled C and F in Fig.\,\ref{MIRA_GRAVITY}) stretching along the northeast direction, which appear to be nearly perpendicular to the axis defining the brighter clumps. When we look at the reconstructed image from SQUEEZE, we see that despite being generally similar to its MIRA counterpart, the structure in the SQUEEZE image is a lot more continuous, somewhat resembling a clumpy disjointed ring seen at an intermediate inclination. The features that correspond to regions A, B, D, E, and G in Fig.\,\ref{MIRA_GRAVITY}, are present in Fig.\,\ref{SQUEEZE_GRAVITY }, however, features C and F are not clearly distinguishable in the SQUEEZE image. This could mean that the linear features are imaging artifacts.

Our highest-resolution (0.1 mas) $K$-band images reveal a very clumpy morphology that surrounds the central elliptical structure (see Fig.\,\ref{MIRA_GRAVITY}). The major clumps seem to almost trace out a ring that has a size of about 5\,mas, while the inner source spans 2\,mas. 
%Additionally, in the 0.1 mas resolution image, we also see two linear features that are perpendicular to the orientation of the clumpy disk. 
It is worth noting that at the adopted pixel size of 0.1\,mas, the source is already super-resolved when compared to the actual achievable resolution with the longest baseline at the VLTI (i.e. $\sim$1\,mas). Thus, the finer details such as the clumps and linear features could be simply image artefacts that arise as a consequence of using too high a resolution. To further explore the effects of super-resolution, we performed image reconstruction using larger pixel sizes of 0.5 and 0.8\,mas. These results can be seen in Figs.\,\ref{MIRA-0.2}--\ref{MIRAk-0.8}. The resulting less super-resolved images show only a single extended feature, whereas the inner feature seen in the more resolved images ceases to be distinguishable. Furthermore, the clumps which were prominent in the 0.1\,mas resolution image also seem to have vanished, however, the extended structure in the 0.5 mas and 0.8 mas images seems to suggest that the circumstellar environment is asymmetrical and with a morphology that could be described as bipolar. The orientation of this bipolar structure is along the same axis defined by the PA from the geometrical model fitting. It is likely that the outer structure is indeed an example of outflows that are expected to be produced in post-merger environments, similar to what is seen in the sub-mm regime in the stellar merger remnant V4332\,Sgr \citep{2018A&A...617A.129K}.

\subsection{$H$ band image reconstruction}\label{H_band_imaging}

\begin{figure}%[hbt!]
    \centering
    \includegraphics[trim=5 0 45 30, clip, width=\columnwidth]{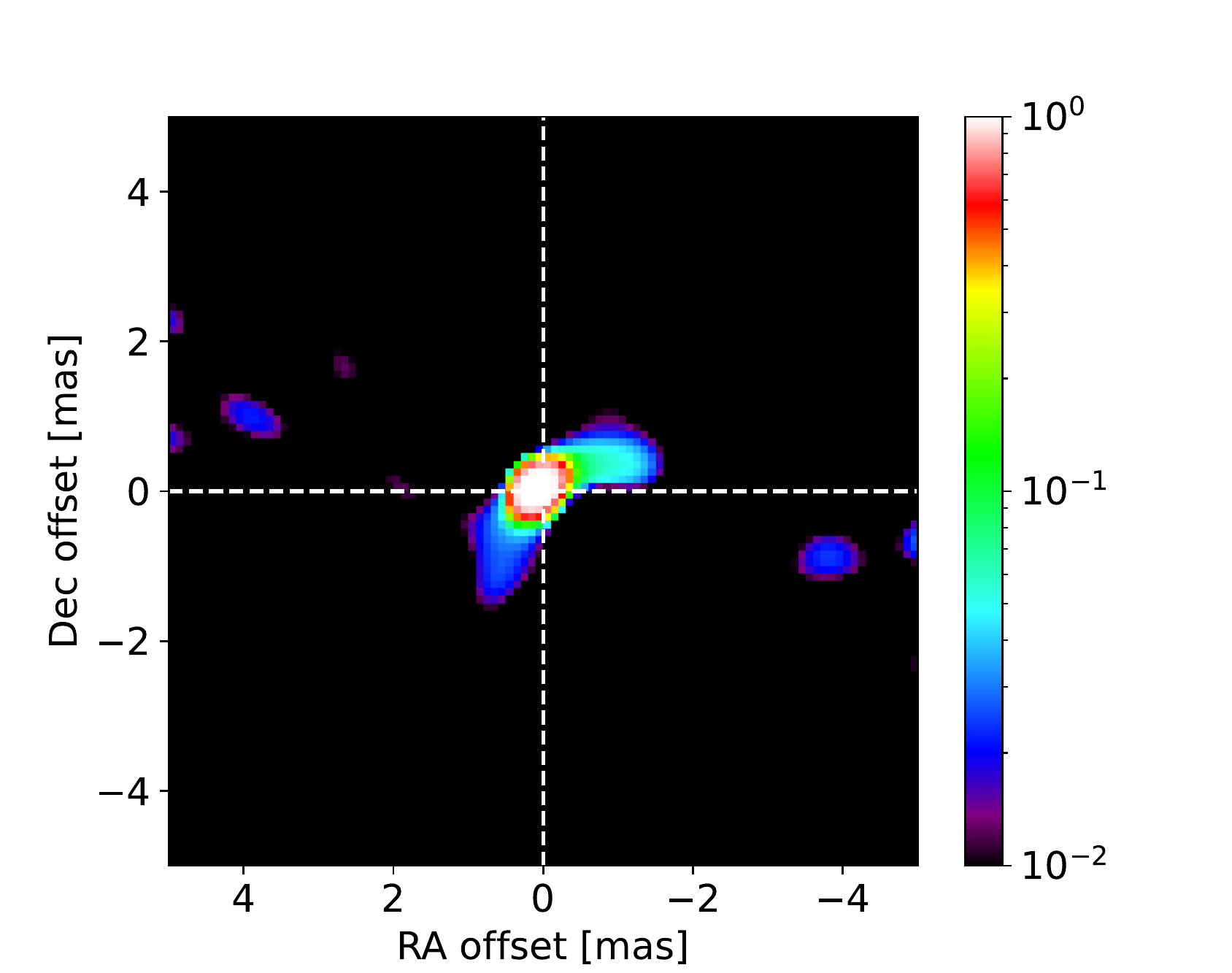}
    \caption{Reconstructed $H$ band image of V838 Mon obtained using the MIRA imaging algorithm.}
    \label{chara_MIRA}
\end{figure}

\begin{figure}%[hbt!]
    \centering
    \includegraphics[trim=5 0 45 30, clip, width=\columnwidth]{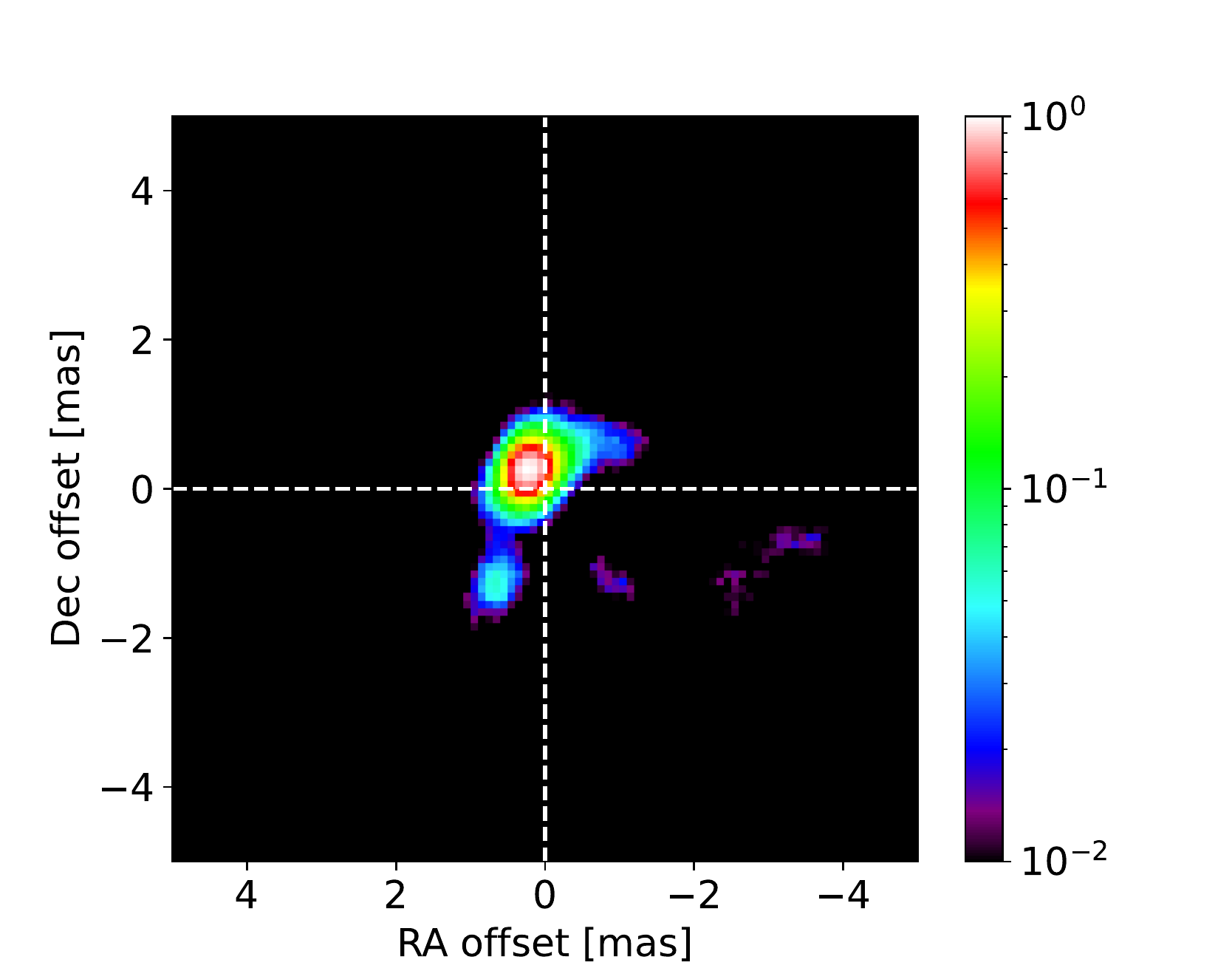}
    \caption{Reconstructed $H$ band image of V838 Mon obtained using the SQUEEZE imaging algorithm.}
    \label{chara_SQUEEZE }
\end{figure}

Similar to how we analyzed the $K$ band data, we employed the same imaging algorithms and attempted image reconstruction for the CHARA observations in the $H$ band. The image FoV was set to 10 mas, while a pixel size of 0.1 mas was chosen and the starting image was randomly generated. The value of the regularization was set to 1000 for both MIRA and SQUEEZE imaging. The resulting images (Figs. \ref{chara_MIRA} and \ref{chara_SQUEEZE }) show an elongated structure that is oriented at a PA of --40\degr. The physical extent of this feature is about 2 mas, thus allowing us to probe the innermost parts of the circumstellar environment. Despite the fact that the images provide a very good fit to the data (i.e. reduced $\chi^{2} = 1$) some artifacts are also present. These might have occurred due to the fact that we were only able to use 4--5 telescopes on any given night and not the full six telescope configuration of CHARA.    

\begin{figure}%[hbt!]
    \centering
    \includegraphics[clip, width=\columnwidth]{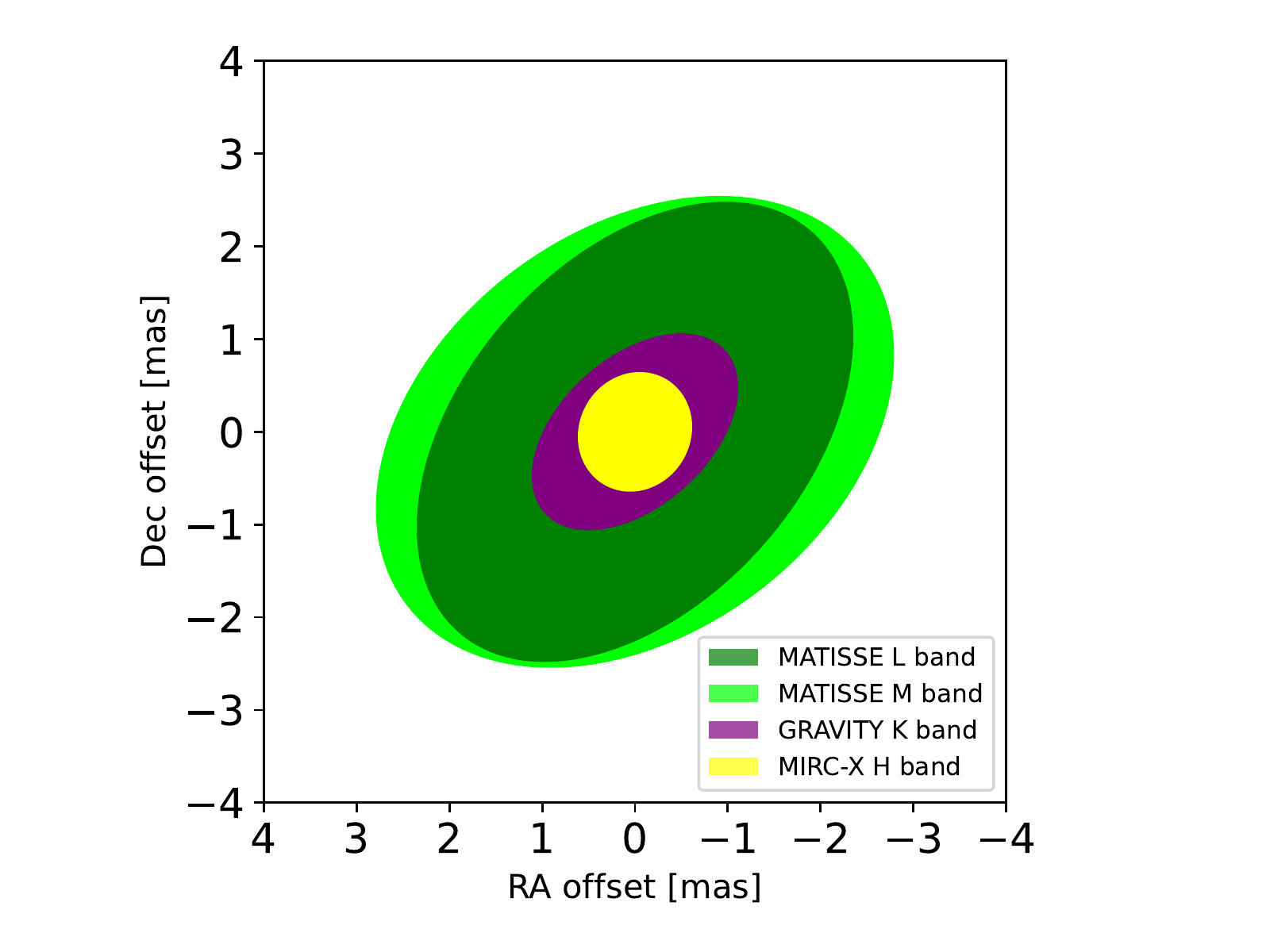}
    \caption{A sketch of the best fitting elliptical models for V838 Mon in the $HKL$ bands as observed by MIRC-X, GRAVITY and MATISSE.}
    \label{fig-cartoon}
\end{figure}

It would seem that in the $H$ band too we are seeing a similar bipolar feature but at much smaller spatial scales, placing it closer to the central merger remnant.
The reconstructed images in the $H$ band show a structure that is similar in morphology to its counterparts in other wavelength regimes. The images indicate the presence of asymmetrical structures in the south-east and north-west regions of V838 Mon. The asymmetry is most noticeable in the northern part of the bipolar structure, which seems to be misaligned with respect to the southern feature. A crucial difference in the $H$ band structure is the relatively high closure phase deviations and the apparent asymmetrical shape of the northern lobe visible in the image reconstructions. The orientation of the CHARA feature is in good agreement with what is seen in the $L$, $M$, and $K$ bands. This can be seen in the sketch of the mid-infrared structures in Fig. \ref{fig-cartoon}.

\subsection{$H$ band closure phase modelling}

\begin{figure}%[hbt!]
    \centering
    \includegraphics[clip, width=0.8\columnwidth]{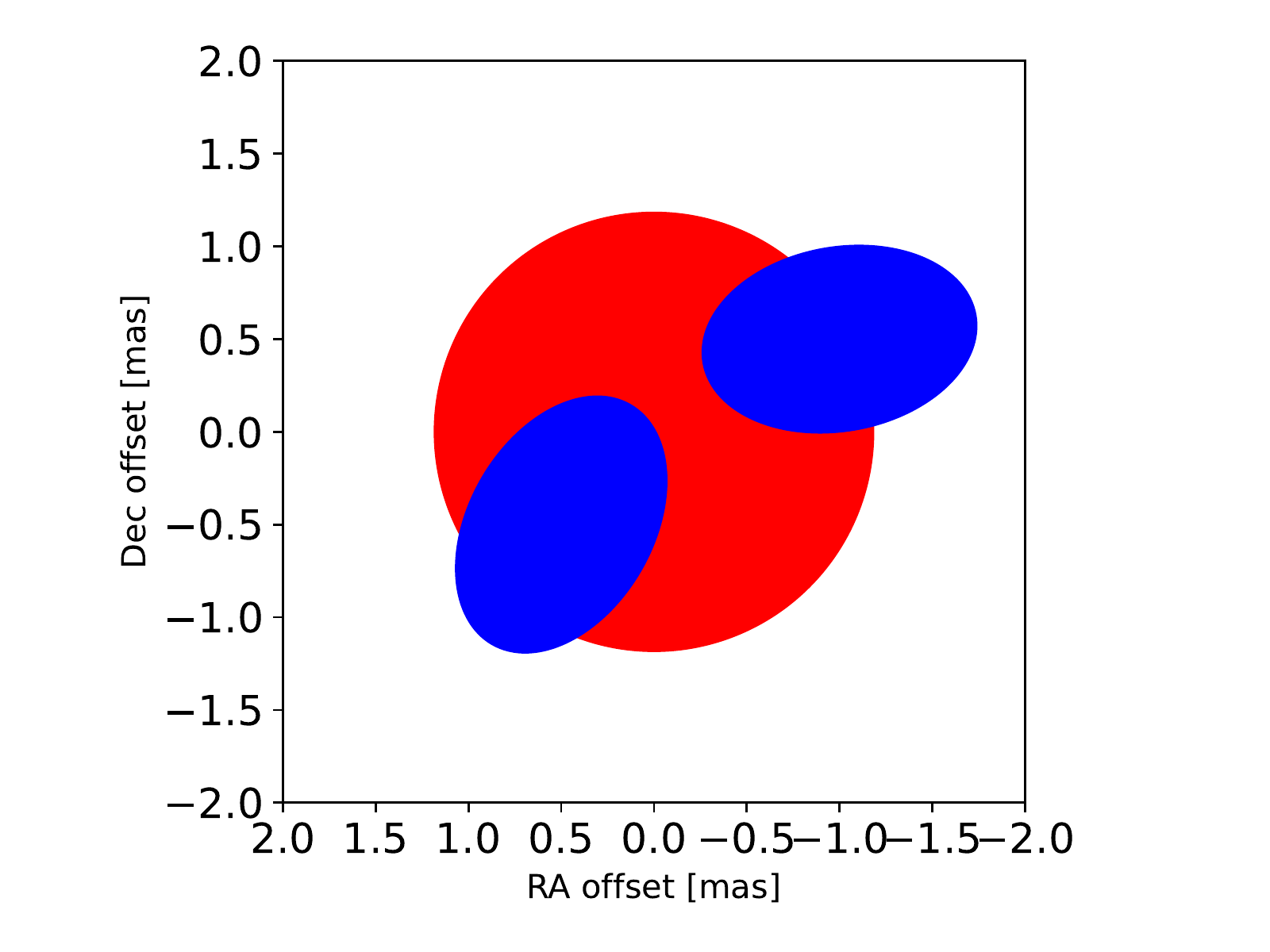}
    \caption{Model we used for explaining the $H$ band closure phases. The red central component represents the star, while the blue ellipses may represent bipolar outflows hinted by the $H$ band image reconstructions (cf. Figs. \ref{chara_MIRA} and \ref{chara_SQUEEZE }). About 95\% of the model flux is contained within the central component. }
    \label{Hband-cartoon}
\end{figure}

\begin{figure}%[hbt!]
    \centering
    \includegraphics[trim=5 5 55 40, clip, width=\columnwidth]{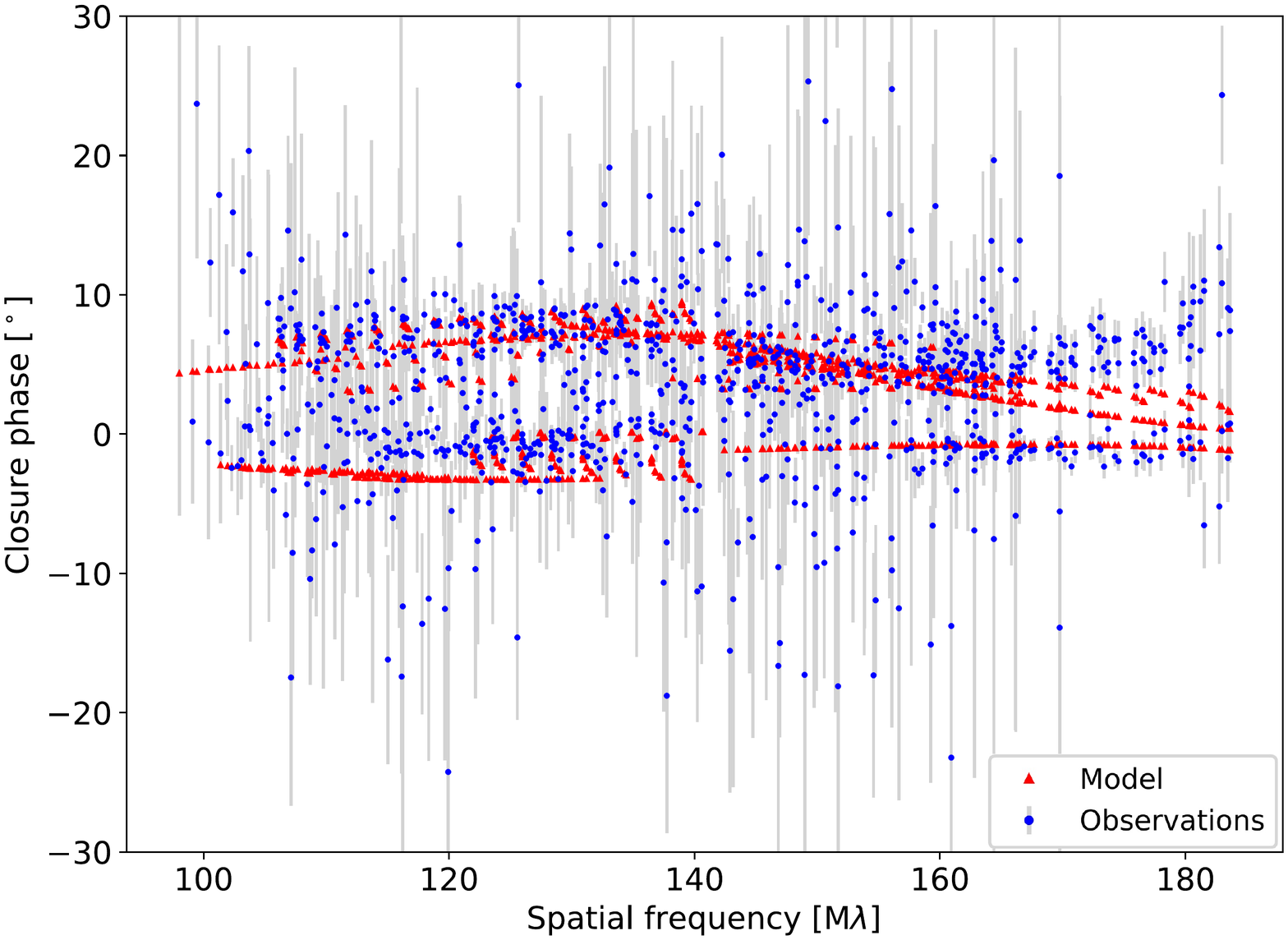}
    \caption{Simulated and observed $H$ band closure phases as a function of spatial frequency. The simulated closure phases are represented by red triangles, while the observations are represented by the blue circles. Simulations correspond to the structure shown in Fig. \ref{Hband-cartoon} and match the observations better than the simpler models described in Sect. \ref{sec-Hgeo}.}
    \label{Hband-cartoon-cps }
\end{figure}

As mentioned previously, we were unable to explain $H$ band closure phases with simple geometrical models (circular and elliptical disks, see Sect.\,\ref{sec-Hgeo}). Given the magnitude of these closure phase deviations, it is apparent that the circumstellar structure in V838\,Mon possesses significant asymmetries. This is further corroborated by our imaging results presented in Sect.\,\ref{H_band_imaging}. Similar to the $K$ band images, the $H$ band ones seem to show a bipolar structure which is roughly oriented along the same PA as all the other structures that encase the merger remnant. Given our imaging results in the $H$ band, we again attempt to reproduce the closure phase deviations, this time using a more complicated multi-component model that better represents the structure as revealed by the $H$ band image. A sketch of this model is shown in Fig. \ref{Hband-cartoon}. The model includes a central 1\,mas disk that represents the central star. In addition to that, we also include two elliptical disks (``clumps"). The semi-major axis of the lower ellipse is oriented roughly along the PA that the structures in other bands follow, while that of the upper ellipse is nearly horizontal. The modelled and observed closure phases are plotted as a function of spatial frequency in Fig. \ref{Hband-cartoon-cps }. We see that the more complex model inspired by the image reconstructions is able to explain the closure phase deviations quite satisfactorily. 

\subsection{MYSTIC $K$ band closure phase modelling} \label{MYSTIC_CPS}

\begin{figure}%[hbt!]
    \centering
    \includegraphics[clip, width=\columnwidth]{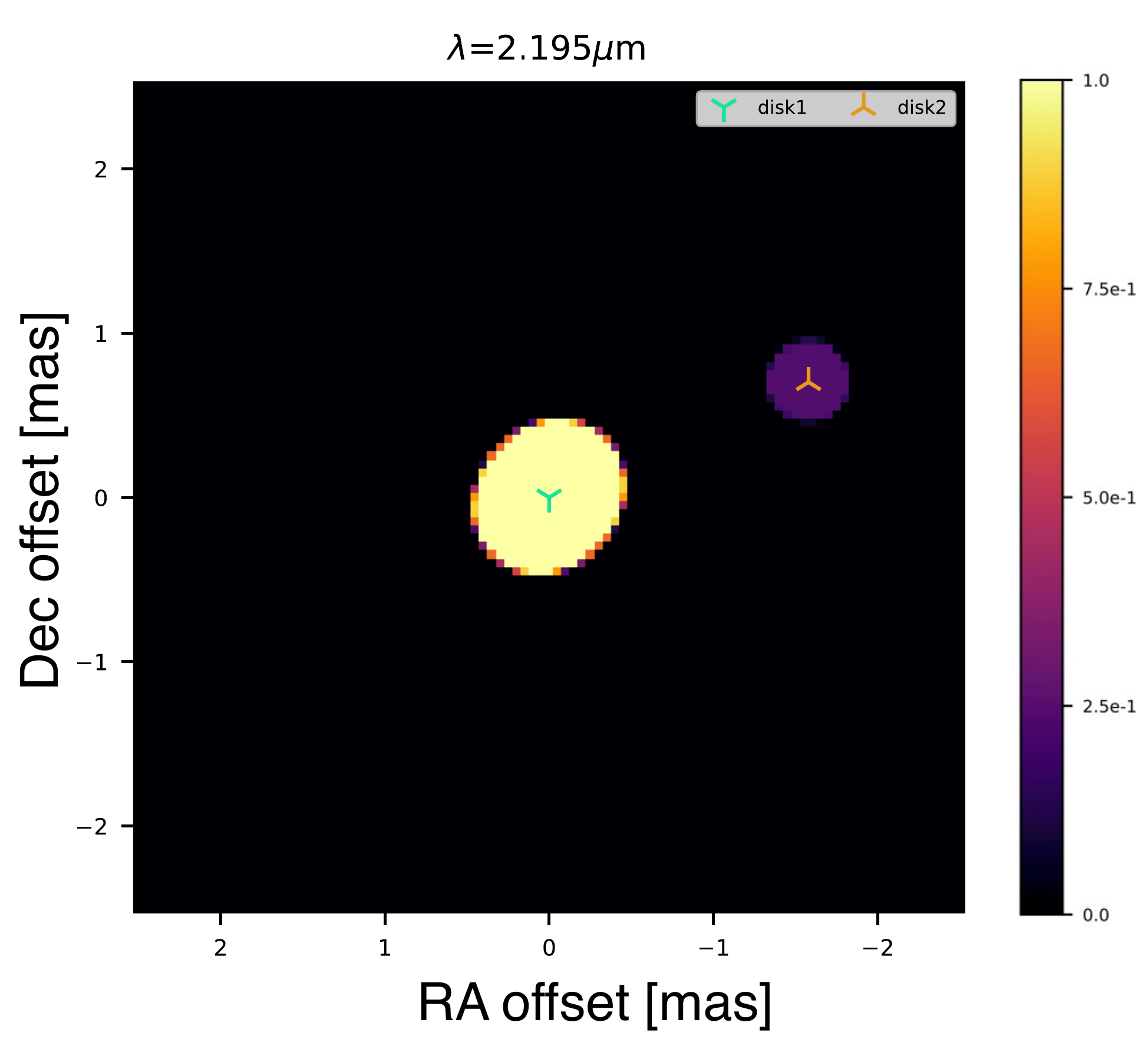}
    \caption{Image for the best-fit geometric model comprising two components in the $K$ band. The primary component is the elliptical disk with a size of $\sim$1\,mas and elongated at a PA of $-40\degr$. The secondary component is a smaller circular disk with a diameter of 0.5\,mas and separated from the primary by $\sim$1.7\,mas. The bulk of the model flux ($\sim$ 99\%) is contained in the primary component.}
    \label{MYSTIC-model}
\end{figure}

As stated before, the $K$ band closure phases as observed by MYSTIC do show significant non-zero deviations. Therefore, we decided to model these closure phases geometrically to see if the observations can be well explained with an asymmetric intensity distribution. We used the interferometric modelling software PMOIRED \citep{2022SPIE12183E..1NM} for this purpose. Using PMOIRED, we experimented with various combinations of geometrical building blocks (disks, Gaussians, etc.) in order to reproduce the closure phases as accurately as possible. Ultimately, we found the best fit model ($\chi^{2}\!\sim$1.5) which consists of an elongated primary disk with a size of 1\,mas and a smaller (0.5\,mas) less intense secondary disk that is separated from the primary by $\sim$1.7\,mas. The geometry of this model is shown in Fig.\,\ref{MYSTIC-model}.   

\begin{figure}%[hbt!]
    \centering
    \includegraphics[clip, width=\columnwidth]{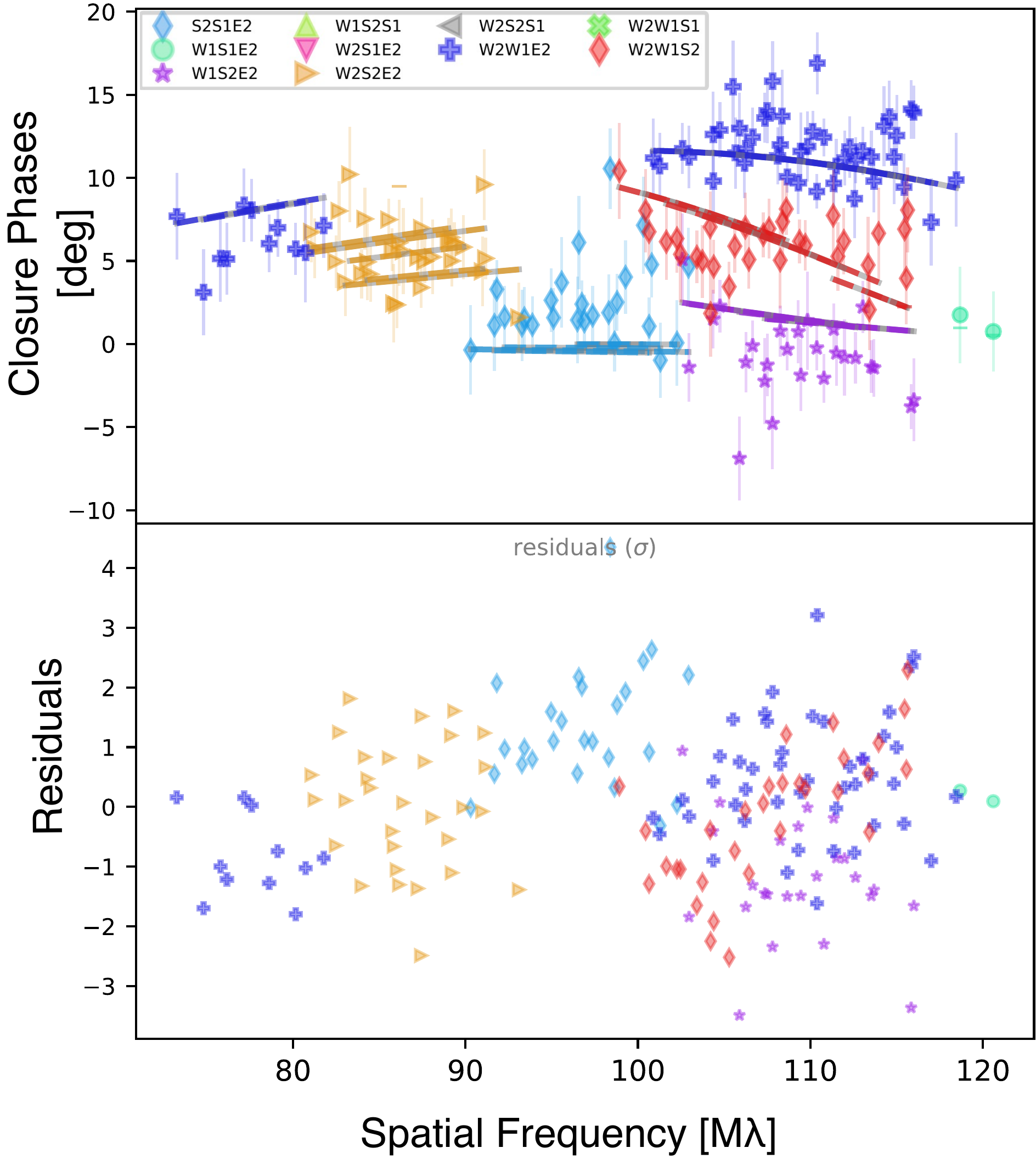}
%    \caption{The UV coverage and baselines for the MYSTIC observations are shown on the left plot, while on the right the fits to the observations along with the residuals are shown.}
    \caption{MYSTIC $K$-band closure phases (top) for the observed (points) and modelled (lines) visibilities representing the model in Fig.\,\ref{MYSTIC-model}.  The bottom panel shows the respective residuals.}
    \label{MYSTIC-fits}
\end{figure}

The closure phases produced by the model and their fit to the observations are shown in Fig.\,\ref{MYSTIC-fits}. We can see from the figure that the non-zero closure phases are explained quite satisfactorily by our model at all baselines. The orientation of the primary disk seems to agree well with the general orientation of V838 Mon as seen in the $HK$ band image reconstructions. Furthermore, the PA of the smaller disk component with respect to the primary also closely resembles the general orientation of V838 Mon in the near to mid infrared bands. The modelling results prove that the closure phase deviations are well explained with the addition of an off-centre ``clump".    

\section{Discussion} \label{disc}

\subsection{Jets}
\cite{2005ApJ...622L.137L} suggested in their study that V838 Mon could potentially possess an asymmetric structure, however due to their limited observations and absence of closure phase measurements they were unable to prove this conclusively. In \cite{2021A&A...655A.100M}, the $L$ band closure phases seemed to be hinting towards minor asymmetries, as indicated by the small but non-zero deviations. However, we interpreted them with great caution given their small magnitude and the limited data. 
In this study, we have conclusively shown the existence of these asymmetries via geometrical modelling and image reconstruction in the $HK$ bands. The GRAVITY and CHARA images (see Figs. \ref{MIRA_GRAVITY} and \ref{chara_MIRA}) show an inhomogeneous bipolar structure. This inhomogeneous structure is more pronounced in the $H$ band image, as we can see that the northern lobe is oriented at a different PA than its southern counterpart. The asymmetric lobes model is further corroborated by $H$ band closure phases that are satisfactorily explained when we assume such a morphology for our geometrical model (see Figs. \ref{Hband-cartoon} and \ref{Hband-cartoon-cps }). 

These outflows resemble jets, which are strongly believed to be present in Intermediate Luminosity Optical Transients (ILOT) such as Luminous Red Novae (LRN) and are thought to play an important role by providing the extra energy needed to unbind the common envelope. \cite{SOKER201616} shows that energy from the jets is transferred to the primary envelope during the grazing envelope evolution stage, which means that the jets could be the mechanism responsible for the outbursts seen in LRNs including V838 Mon. In a subsequent study, \citep{2020ApJ...893...20S} it was shown that the luminosity and total energy output of V838 Mon could be adequately explained if jets present in the system collide with a slow moving shell of material. The author concluded that this was a plausible scenario for V838 Mon and V4332 Sgr. In the case of V4332 Sgr it was shown using sub millimetre observations carried out at ALMA \citep{2018A&A...617A.129K} that the object does indeed possess bipolar outflows consisting of molecular material. This means that jets were a main component in the LRN event in V4332 Sgr. If jets are the reason for the LRN in V838 Mon, then that necessitates that we should be able to observe jet like features in the source. Since we see a clear bipolar structure in the image reconstructions for V838 Mon presented in Sects. \ref{K_band_imaging} and \ref{H_band_imaging} this means that jets played a role in the energy output of the outburst in 2002 which resulted in V838 Mon brightening by several orders of magnitude. \cite{2023arXiv230607702S} shows that only jets can satisfactorily explain the rapid rise in luminosity that is seen in ILOTS such as V838 Mon and V1309 Sco just prior to the outbursts. As of now, no jets have been observed in V1309 Sco, but the clear detection of jets in V838 Mon, V4332 Sgr and CK Vulpeculae suggests that such bipolar features are perhaps universal amongst many other stellar merger remnants. Recently the Blue Ring Nebula was found to possess bipolar conical outflows which led \cite{2020Natur.587..387H} to conclude that the object is a result of a stellar merger. Our interferometric observations of V838 Mon seem to be giving credence to the idea that jets are an intrinsic feature of LRNs and persist even in the post-merger environment. In V838 Mon it seems that the jets are only prominent in the mid-infrared $HK$ bands since the ALMA maps show the merger ejecta to be spherically symmetric at a spatial scale of $\sim$ 25 mas and any jet-like feature seems to be absent \citep{alma}. This contrasts with the ALMA maps for V4332 Sgr which show a very clear signature of a bipolar outflow \citep{2018A&A...617A.129K}. With extensive future VLTI observations of V4332 Sgr we will be able to determine the extent to which the milliarcsecond scale environment differs from the large scale structures as revealed by ALMA. It could be that the nature of every LRN jet is unique and is determined by a number of other factors such as the mass and stellar type of the progenitor stars, the surrounding medium and duration of the outburst, etc.     

\subsection{Clumps}

The super resolved GRAVITY $K$ band image (see Fig. \ref{MIRA_GRAVITY}) points to the possible presence of clumps in the immediate circumstellar vicinity in V838 Mon. The GRAVITY closure phases alone do not seem to give us much insight, as the deviations are much too small (see lower panel in Fig. \ref{Fig-Kvis+phases}). However, since the CHARA baselines are longer by a factor of about three, the closure phases measured using MYSTIC show much higher deviations from $0 \degr$ (discussed in Sect. \ref{MYSTIC_CPS}). These MYSTIC closure phases, as evidenced by the modelling, suggest the presence of a small clump separated from the central star by $\sim 1.7$ mas. The size of the clump is 0.5 mas, which makes it too small to be detected by the VLTI but large enough to appear in our CHARA observations. 

The clump lies approximately along the PA traced by the bipolar structure in V838 Mon, in fact its orientation seems to be quite similar to that of the northern lobe seen in the $H$ band. This clump is the first ever detection of a sub milliarcsecond feature in V838 Mon. The exact nature of the clump, whether stable or transient as of now, remains elusive. ALMA maps of V838 Mon by \cite{alma} also showed the presence of prominent clumps in the wind component of the merger remnant in the submillimeter regime. The formation of dust clumps is a prevalent phenomenon amongst red supergiants (RSGs) and asymptotic giant branch stars. VX Sagitarri which is a variable supergiant star has been found to possess an inhomogeneous atmosphere which contains spots \citep{2010A&A...511A..51C}. It was shown by \cite{2009A&A...506.1351C} using radiative hydrodynamical simulations that the interferometric observations for the RSG $\alpha$ Ori could be explained using limb darkening models, which suggested the presence of asymmetric surface features. In addition to this, \cite{2019A&A...627A.114K} showed using ALMA observations that multiple dust clumps had formed in the RSG VY Canis Majoris (VY CMa). A dust clump is also thought to have formed in the (RSG) Betelgeuse during the Great Dimming event of 2019--2020 \citep{2021Natur.594..365M}. Follow-up VLTI/MATISSE observations in 2020 were obtained by \cite{2023arXiv230308892C} which showed a complex closure phase signal in Betelgeuse. The authors constructed multiple radiative transfer models, such as those consisting of clumps and spots. Although they could not reproduce the closure phase signal exactly due to their use of only rudimentary models, they concluded that either clumps or other surface features were the main source of the closure phase deviations. In the case of V838 Mon, the closure phases are not as high as those in Betelgeuse observed by \cite{2023arXiv230308892C}, but, as shown in Sect. \ref{MYSTIC_CPS}, the MYSTIC closure phases can only be explained by the smaller clump (see Fig. \ref{MYSTIC-model}). 

Since the spectra for V838 Mon resembles that of a late M-type supergiant, \citep{2003MNRAS.343.1054E,2015AJ....149...17L} we expect the circumstellar environment in V838 Mon to be somewhat similar to other evolved stars with clumpy surroundings. This particular clump that we have been able to constrain with MYSTIC could in fact be a smaller feature within the lobe/jet like structure that is observed at longer wavelengths. This would mean that the lobes themselves could be clumpy and inhomogeneous. These clumps could be the result of the jet interacting with the surrounding less dense material in the V838 Mon during its common envelope phase. \cite{2019MNRAS.483.5020S} propose that this interaction between the jet and the envelope results in the formation of jet-inflated low-density bubbles. As a consequence, the surrounding medium is prone to clumping due to Rayleigh-Taylor instabilities. This would result in the jet departing from a pure axisymmetric morphology and become variable in direction and brightness. It could be the case that the observed $H$ band clump was the result of such jet-inflated bubbles.

If the clump is a transient feature, then multiple follow-up observations with MIRCX/MYSTIC at CHARA could help to trace out its trajectory with respect to the central merger product. These observations could also help determine the time evolution of the size and brightness of the clump, which can help to determine whether the jet-inflated bubble formation scenario is plausible. 

\subsection{Clump ejection scenario}

The outburst in 2002 showed a complex multi-peaked light curve that was explained to be the result of multiple shell ejection events in V838 Mon. These short duration bursts of energy are thought to coincide with the ejection of the shells and material outflows \citep{2005A&A...436.1009T,2007ASPC..363..280S}. Recent $V$ band photometry of V838 Mon also shows that the source has been gradually increasing in brightness over the past decade, especially in the $UBV$ bands \citep{2023A&A...670A..13L}.
The photometry also shows amplitude variations of about $\sim$0.4, which occur with an approximate period of $\sim$300 days \footnote{\url{https://asas-sn.osu.edu}}. 

If the observed clump is a mass ejection event or part of the bipolar outflows seen in the $H$ band images (see Fig. \ref{chara_MIRA}), then it could very likely be the case that the formation of such features correlates with the small brightness increases that are seen in the light curve modulations for V838 Mon. If true, then this would mean that the post-merger circumstellar environment is quite dynamic and is evolving over time scales of about a year. This is in stark contrast to the $L$ band structure that appears to be stable and non-transient over timescales of years. Future extensive observations of V838 Mon with MYSTIC/MIRC-X could help to determine if the observed amplitude variations correlate with the appearance of 'clumps', which can help to constrain the source of the brightness fluctuations observed in V838 Mon.

 \subsection{Contraction controversy}
 
 Previously, it was claimed that V838 Mon was gradually shrinking due to the geometrical thinning of the surrounding ejecta \citep{2014A&A...569L...3C}. Our recent size measurement in the $K$ band shows that contraction has apparently "ceased" and the size of V838 Mon at 2 $\mu$m is now greater than that measured by \cite{2005ApJ...622L.137L} just a few years after the outburst. This would imply that V838 Mon instead has now started expanding since the previous size measurement by \cite{2014A&A...569L...3C}. Previous interferometric observations by \cite{2005ApJ...622L.137L} and \cite{2014A&A...569L...3C} were limited in scope due to a wide variety of factors. 
 
 For example, the observations by \cite{2005ApJ...622L.137L} were done using the PTI which had only two baselines with a maximum baseline length of 85 m. As a result of this, V838 Mon was not properly resolved, as is evidenced by the fact that the lowest squared visibility measured by \cite{2005ApJ...622L.137L} was about 0.6. Similarly, the VLTI-AMBER squared visibilities at 2 $\mu$m fitted by \cite{2014A&A...569L...3C} reached a minimum of about 0.2, which is a significant improvement over the previous measurement, but a small fraction of the source flux remained unresolved. This also explains why the size measurement obtained by \cite{2014A&A...569L...3C} is smaller than that estimated by \cite{2005ApJ...622L.137L}. Their fit yielded a size of 1.15 $\pm$ 0.20 mas, which is about $40 \%$ less than the value measured by \cite{2005ApJ...622L.137L} (1.83 $\pm$ 0.06 mas), which seemed to have indicated that a contraction had taken place over the course of a decade. 
 
 Our GRAVITY and MYSTIC $K$ band observations on the other hand were able to better resolve V838 Mon due to the longer baselines (maximum baselines of $\sim$ 130 m and $\sim$ 300 m on the VLTI and CHARA respectively) and greater instrument sensitivity. The smallest observed value for $V^{2}$ with VLTI was about 0.1 while with CHARA it was $\sim$ zero. These size estimates for V838 Mon are based on observations that almost fully resolve V838 Mon, which was not the case in the above-mentioned previous studies. For example the uniform disk diameter measured with CHARA-MYSTIC as stated earlier is $1.57 \pm 0.01$ mas while with VLTI-GRAVITY it is $1.939 \pm 0.001$ mas (see Table \ref{model params}), since the CHARA baselines are longer than the VLTI ones by about a factor of three we conclude that the CHARA size estimate is indeed the most accurate. If the initial size estimate by \cite{2005ApJ...622L.137L} is assumed to be accurate, then our results imply that indeed V838 Mon has begun to expand. However, as we have discussed above, the 2005 measurements were limited by baseline length, and therefore it is highly likely that the angular diameter for V838 Mon was overestimated. This would explain the apparent contraction that was believed to be occurring in V838 Mon. Future observations with CHARA can help to determine the actual evolution of the size of V838 Mon by comparing them to the previous measurement from 2013 and the ones presented in this study.

 %A possible reason for this reversal in contraction could be the accretion of some of the surrounding ejecta material onto the newly formed merged star. This fallback accretion could then result in the production of short-lived jets that originate from the central star, similar to how jets arise due to accretion of material onto the secondary in a post-AGB binary system \citep{2022A&A...666A..40B}. 

\subsection{Polarization in V838 Mon}

The Gemini speckle image reconstructions in the $V$ and $I$ band speckle images show a morphology that is consistent with the interferometric imaging and modelling results in the mid-infrared. Both images (see Figs. \ref{gemini-Iband} and \ref{gemini-vband}) seem to suggest that the innermost vicinity in V838 Mon is quite noticeably bright at 562 nm and 832 nm. If this emission results from scattered light, then it is likely to produce a net polarization. It was shown by \cite{wisniewski2003a} that a low degree of polarization (1$\%$) was present in 2003 in the $V$ band, with a measured PA of $150\degr$. The authors suggested that this polarization signal was pointing to asymmetries in the surroundings of V838 Mon. A follow-up measurement, six months later, showed that the PA of the polarisation vector flipped by $90\degr$ \citep{wisniewski2003b}. This change over the course of six months seemed to suggest the dynamic, evolving nature of the dusty environment in V838 Mon. The Gemini observations reveal a complex circumstellar medium in V838 Mon that might still be dynamic and could show signs of varying intrinsic polarization and PA, especially in the $V$ band. The above-mentioned observations by \cite{wisniewski2003a} and \cite{wisniewski2003b} were carried out over two decades ago, and since the dramatic change in PA occurred over a period of only six months, we expect the environment in V838 Mon to have significantly changed since then. Future polarimetric measurements could help to determine if linear polarization in V838 Mon has increased significantly in the last two decades. As we discuss earlier in the section, it could be possible that the observed jet like features in the $K$ band could be transient phenomenon, in which case it would be important to study how the PA and degree of polarization change alongside the jets. We also note that the clumping scenario that we have discussed before could also be probed with further polarimetric observations. For instance, if the clump like feature that we see in our MYSTIC $K$ band modelling results is indeed a mass ejection event, then it could cause significant changes in the degree of polarization and the orientation of the PA as it propagates through the dusty circumstellar environment.

These polarimetric observations can be combined alongside our interferometric results in this paper for radiative transfer modelling of the dusty environment in V838 Mon. Synthetic images at the $HK$ bands can be produced from these simulations, which can then be compared to the reconstructed images that we presented in this study.

%However, further observations are required to ascertain the nature of these jets/lobes in the $H$ band. Multiple follow-up observations in the $H$ band using MIRCX at CHARA could be used to track the evolution of these 'jets' and determine if these are indeed long-lived stable features or a transient phenomenon. 

%Bipolar jets are predicted from numerical simulations of stellar mergers and are found to form in the pre-common envelope stage due to accretion onto the central binary. This bipolar structure has also been observed in the luminous red nova remnant V4332 Sagittari which is thought to have also been the product of a stellar merger.  Our mid-infrared observations of V838 Mon clearly hint towards the presence of such a structure in the $HK$ bands, which seems to be suggesting that there is a possible correlation between the presence of jet-like features and post merger remnants.   

\section{Conclusions} \label{concl}

In this study, we were able to perform a multi-wavelength interferometric analysis of the stellar merger remnant V838 Mon in the $HKLM$ bands using the VLTI instruments MATISSE and GRAVITY as well as the CHARA instrument MIRC-X/MYSTIC. In the $HK$ bands we imaged the circumstellar environment in the immediate vicinity of the central merger remnant, while in the other bands we resorted to geometrical modelling to obtain constraints on the geometry and orientation of the post-merger environment. We were also able to compare the mid-infrared image reconstructions with $I$ and $V$ band speckle imaging from Gemini South. We also modelled the spectral features that were present in the $K$ and $M$ bands and were able to put constrains on the temperatures. Our main findings are as follows.

\begin{itemize}

\item Geometrical modelling suggests that the $L$ band structure has not varied at all since previous observations presented in \cite{2021A&A...655A.100M}.   

\item Image reconstruction in the $HK$ bands reveals a bipolar morphology that resembles jets. These 'jets' are present in both bands, while in the $H$ band they appear to be slightly asymmetric as well. 

\item The super resolved GRAVITY $K$ band images at 0.1 mas reveal clumpy outflows that surround the inner feature. 

\item The MYSTIC $K$ band closure phases are well explained by a small (0.5 mas), off-centre circular feature that we identify as a potential clump.  

\item The orientation of the extended circumstellar feature (i.e. the outflows) tends to be along the same general direction i.e. north-west, with a PA that varies from --30$\degr$  to  --50$\degr$ and an ellipticity in the range 1.1--1.6, depending on the wavelength, which is in agreement with prior studies \citep{2014A&A...569L...3C,alma}. The $N$ and $I$ band structure however is oriented along the north-south direction

\item The $K$ band CO and water features are best modelled at temperatures of 2000 -- 3000 K, thus hinting at their photospheric origin. We were also able to measure the sizes of the CO and water emitting regions, which are 2.1 mas and 1.985 mas respectively. In the $M$ band, we observe CO for the first time at 4.6 $\mu$m for the first time using interferometry.

\end{itemize}

\begin{acknowledgements}
T. K. and M. Z. M. acknowledge funding from grant no 2018/30/E/ST9/00398 from the Polish National Science Center. Based on observations made with ESO telescopes at Paranal observatory under program IDs 0104.D-0101(C) and 0108.D-0628(D). This research has benefited from the help of SUV, the VLTI user support service of the Jean-Marie Mariotti Center (\url{http://www.jmmc.fr/suv.htm}). This research has also made use of the JMMC's  Searchcal, LITpro, OIFitsExplorer, and Aspro services. This work is based upon observations obtained with the Georgia State University Center for High Angular Resolution Astronomy Array at Mount Wilson Observatory.  The CHARA Array is supported by the National Science Foundation under Grant No. AST-1636624 and AST-2034336. Institutional support has been provided from the GSU College of Arts and Sciences and the GSU Office of the Vice President for Research and Economic Development. Time at the CHARA Array was granted through the NOIRLab community access program (NOIRLab PropID: 2022A-426176; PI: M. Mobeen). SK acknowledges funding for MIRC-X from the European Research Council (ERC) under the European Union's Horizon 2020 research and innovation programme (Starting Grant No. 639889 and Consolidated Grant No. 101003096). JDM acknowledges funding for the development of MIRC-X (NASA-XRP NNX16AD43G, NSF-AST 1909165) and MYSTIC (NSF-ATI 1506540, NSF-AST 1909165). Some of the observations in this paper made use of the High-Resolution Imaging instrument Zorro and were obtained under Gemini LLP Proposal Number: GN/S-2021A-LP-105. Zorro was funded by the NASA Exoplanet Exploration Program and built at the NASA Ames Research Center by Steve B. Howell, Nic Scott, Elliott P. Horch, and Emmett Quigley. Zorro was mounted on the Gemini South telescope of the international Gemini Observatory, a program of NSF’s OIR Lab, which is managed by the Association of Universities for Research in Astronomy (AURA) under a cooperative agreement with the National Science Foundation. on behalf of the Gemini partnership: the National Science Foundation (United States), National Research Council (Canada), Agencia Nacional de Investigación y Desarrollo (Chile), Ministerio de Ciencia, Tecnología e Innovación (Argentina), Ministério da Ciência, Tecnologia, Inovações e Comunicações (Brazil), and Korea Astronomy and Space Science Institute (Republic of Korea).

\end{acknowledgements}

\bibliographystyle{aa}
\bibliography{export-bibtex.bib}

\begin{appendix}
\section{Images at lower resolutions}

\begin{figure}%[hbt!]
    \centering
    \includegraphics[clip, width=\columnwidth]{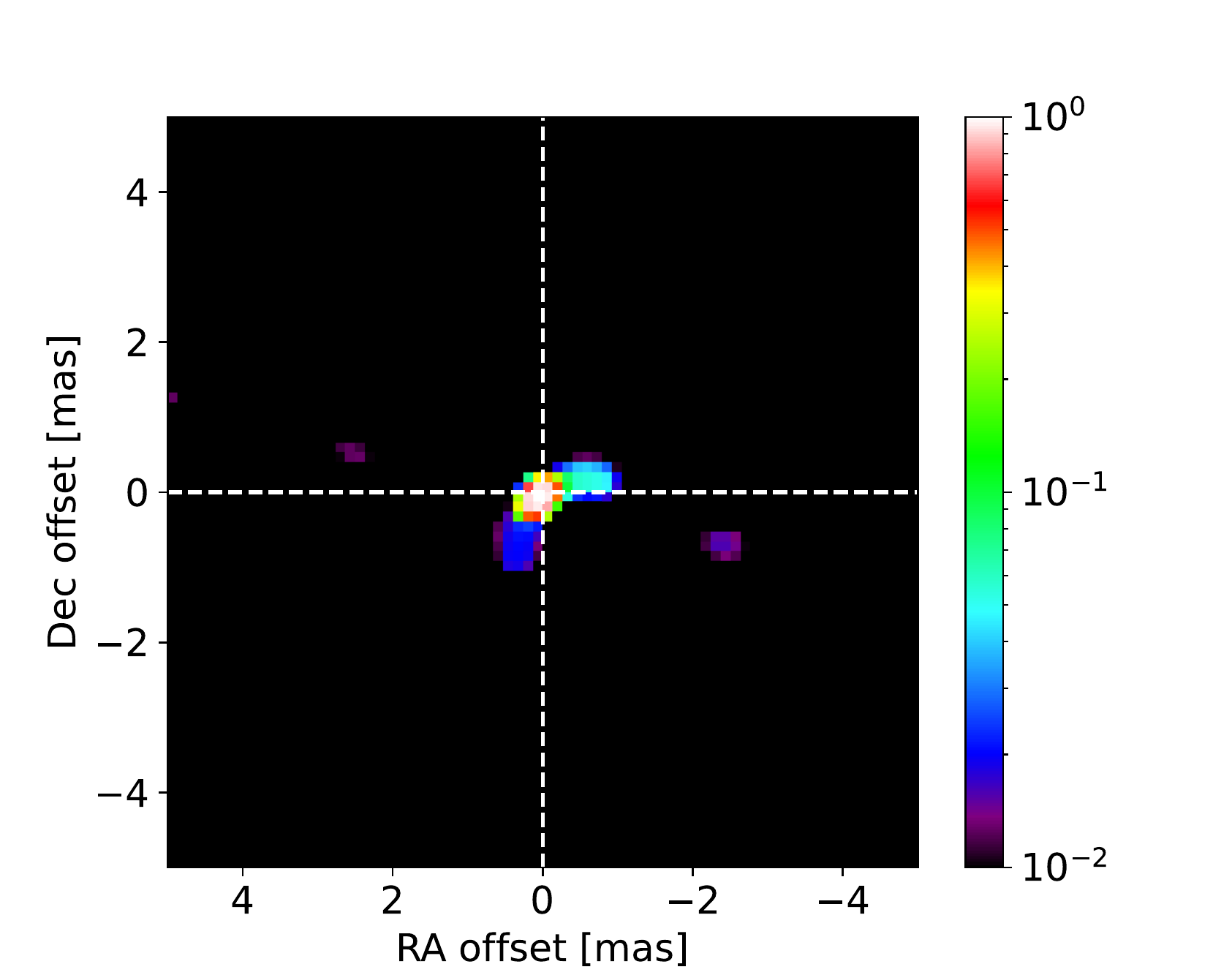}
    \caption{$H$ band image for V838 Mon with MIRA algorithm at a resolution of 0.2 mas.}
    \label{MIRA-0.2}
\end{figure}

\begin{figure}%[hbt!]
    \centering
    \includegraphics[clip, width=\columnwidth]{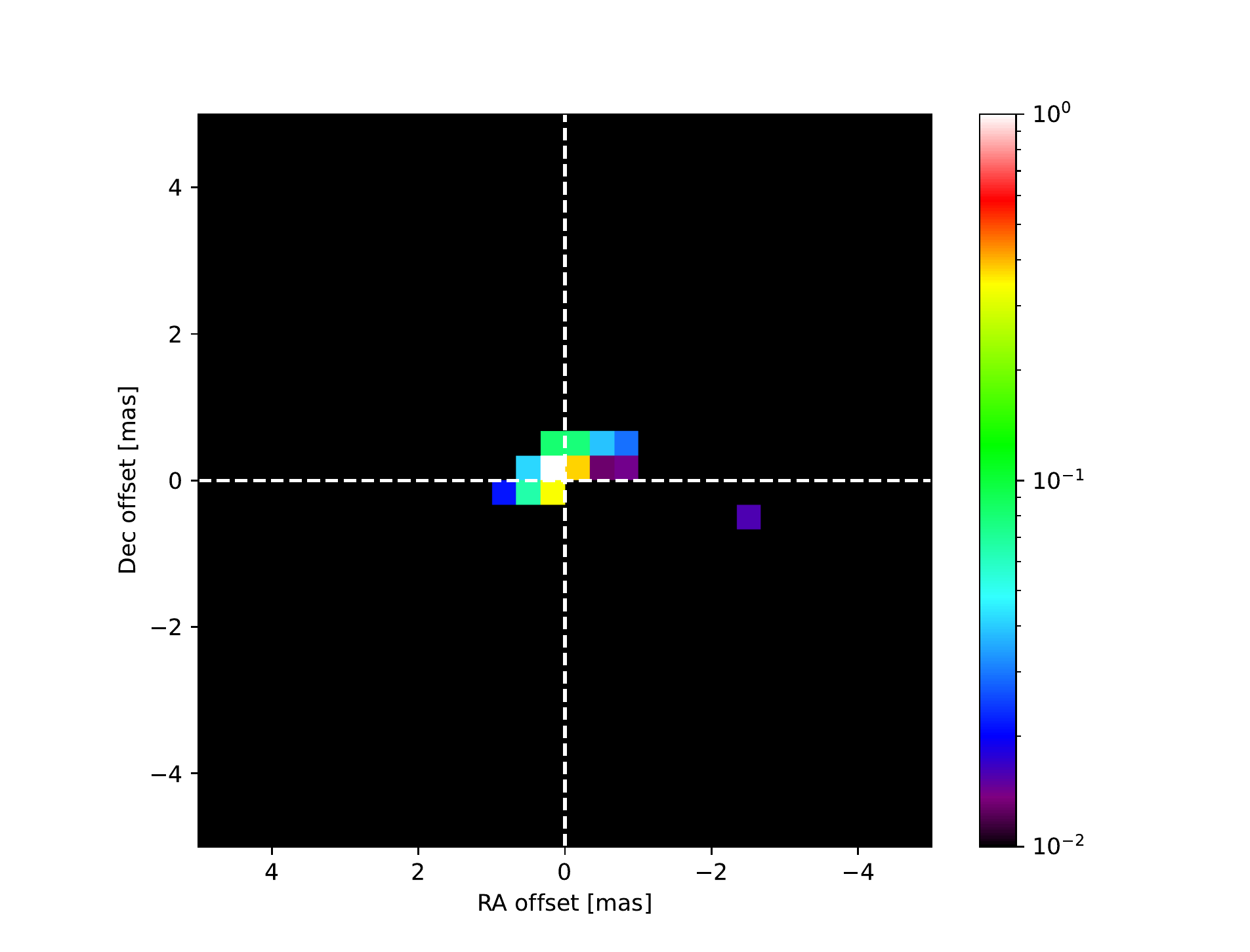}
    \caption{$H$ band image for V838 Mon with MIRA algorithm at a resolution of 0.5 mas. }
    \label{MIRA-0.5}
\end{figure}

\begin{figure}%[hbt!]
    \centering
    \includegraphics[clip, width=\columnwidth]{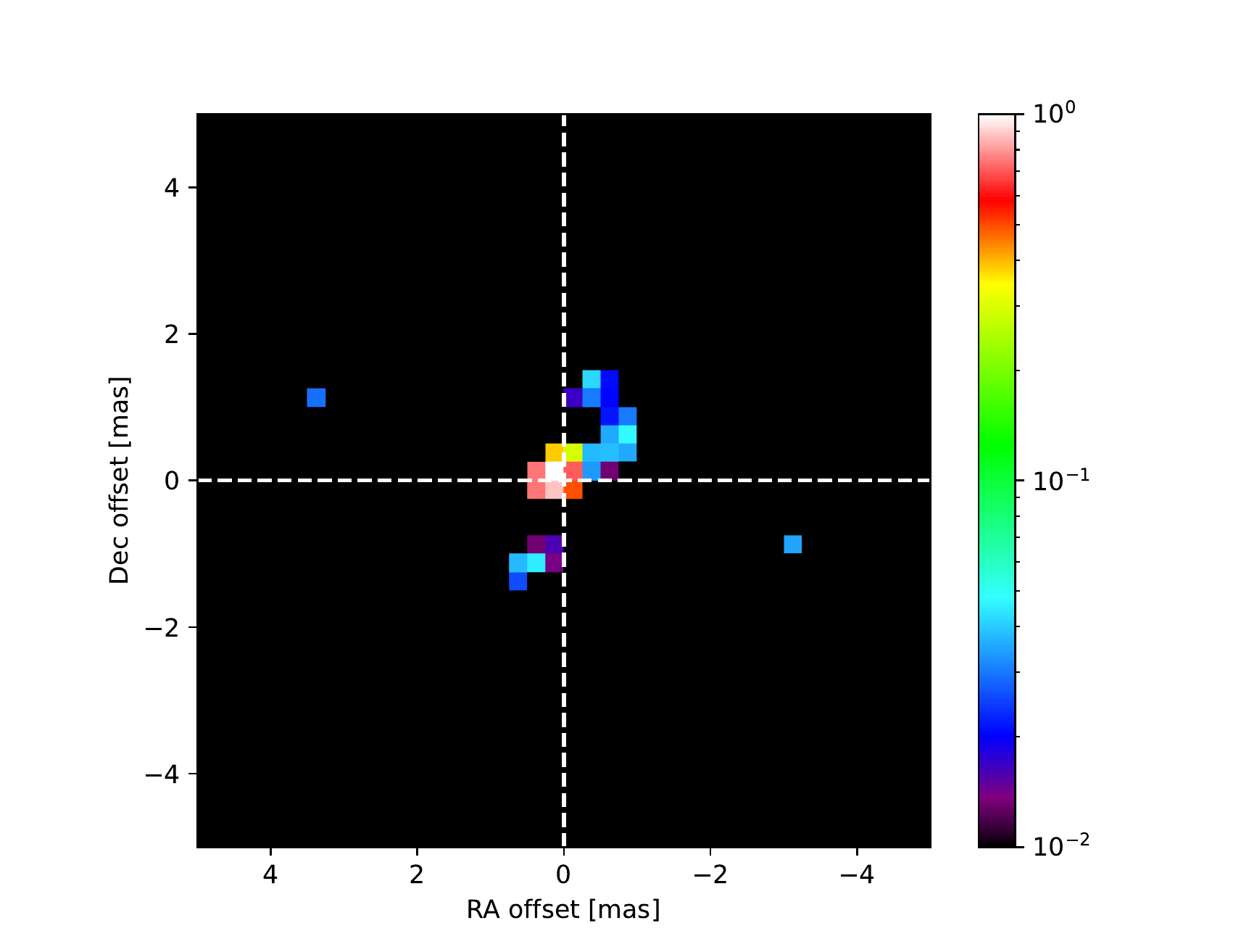}
    \caption{$K$ band image for V838 Mon with MIRA algorithm at a resolution of 0.5 mas. }
    \label{MIRAk-0.5}
\end{figure}

\begin{figure}%[hbt!]
    \centering
    \includegraphics[clip, width=\columnwidth]{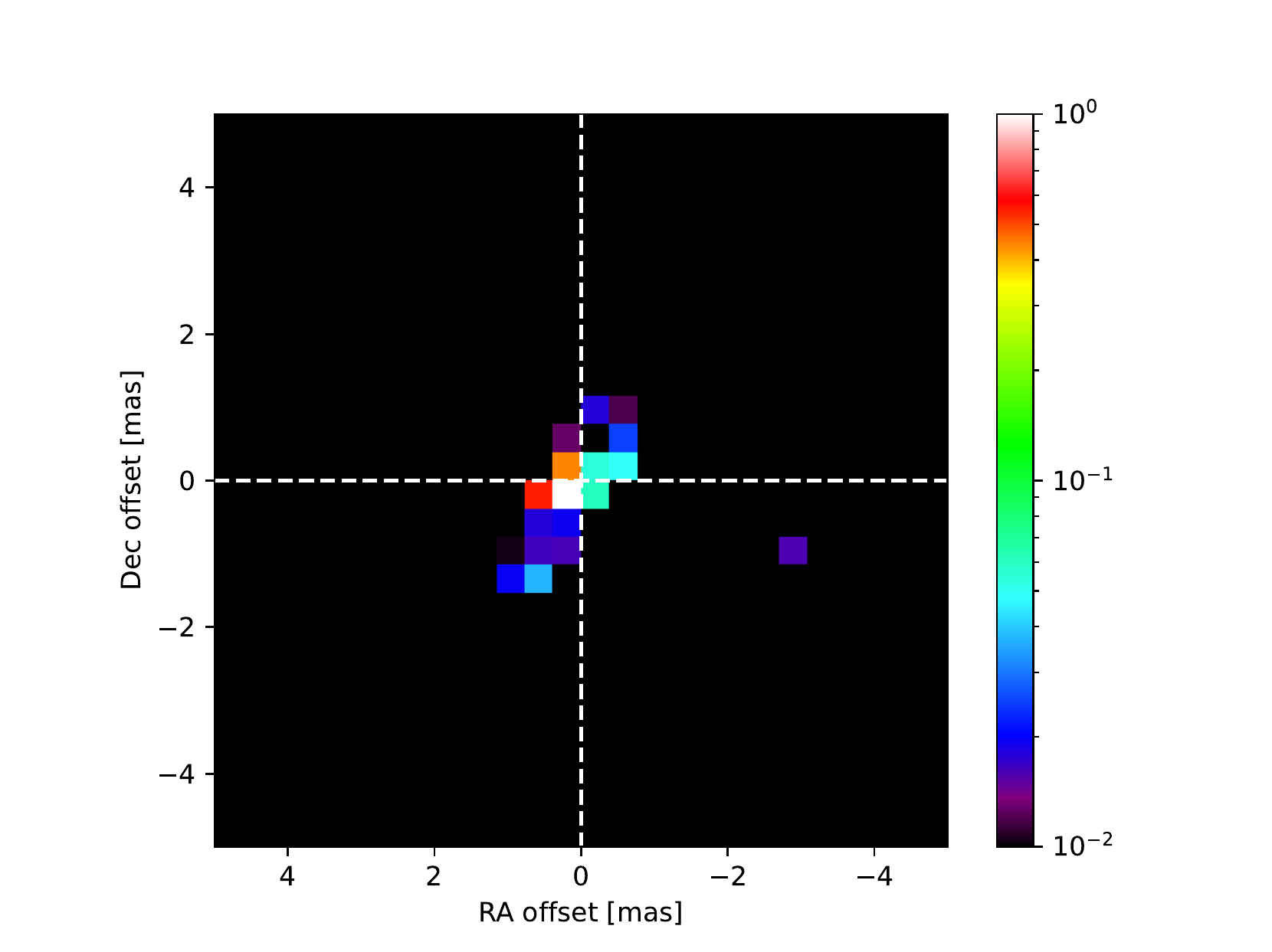}
    \caption{$K$ band image for V838 Mon with MIRA algorithm at a resolution of 0.8 mas. }
    \label{MIRAk-0.8}
\end{figure}

%\begin{SCfigure*}
%    \centering
%    \includegraphics[width=12cm]{modelMatisse.pdf}
%    \caption{Image of the V838\,Mon model at 3.5\,$\mu$m based on the SED and ALMA images. The image is shown in the linear (left) and logarithmic (right) scale. The component seen in the left map north-west from V838\,Mon is related to the B-type companion. For the presentation purposes, the image was smoothed to a resolution of 2\,mas. Colors represent brightness in units of Jy per beam.}
%    \label{fig-almamodel}
%\end{SCfigure*}

\begin{table*}[hb]
\caption{Fitted models and their parameters.}
\centering
%\hspace*{-1cm}
\begin{tabular}{l c c cc  c c c cccc}
\hline\hline
Model & Band & $\chi^2_{\rm r}$ & Size  & $PA$  & Stretch & Flux  \\ 
& & & [mas] & [\degr] &ratio&[\%]& \\
\hline
%[0.5ex] % inserts table %heading
CD  &$L$ &10.34 & 3.00 $\pm$ 0.01 &- & - &$100^{a}$ \\
ED  &$L$ &4.22  & 3.69 $\pm$  0.02 & --41.34 $\pm$ 0.49 & 1.56 $\pm$ 0.01 &$100^{a}$ \\
CG  &$L$ &9.50  &1.82  $\pm$ 0.01 &-  & -& $100^{a}$ \\
EG  &$L$ &4.02 & 1.44 $\pm$ 0.01 &--41.56 $\pm$ 0.51 &1.54 $\pm$ 0.2 &$100^{a}$ \\[5pt]
CD  &$M$ &21.94 & 5.86 $\pm$ 0.08 &- & - &$100^{a}$ \\[5pt] 
EG  &$M$ &18.75 &4.35 $\pm$ 0.09 &--53 $\pm$ 2.87  & 1.42 $\pm$ 0.03  & $100^{a}$ \\
CD  &$H$ &16.92 &1.180 $\pm$ 0.004 & -&-&$100^{a}$\\
ED  &$H$ &16.66 &1.20 $\pm$ 0.01 & --30.69 $\pm$ 3.91 & 1.10 $\pm$ 0.01 & $100^{a}$ \\
CD  &$K$ &341.5 &1.939 $\pm$ 0.001 &-&- &$100^{a}$\\
ED  &$K$ &195.5 &2.619 $\pm$ 0.002 &--47.09 $\pm$ 0.06 &0.615 $\pm$ 0.01 &$100^{a}$\\
CG  &$K$ &328.1 &1.179 $\pm$ 0.001 &-&- &$100^{a}$\\

\hline
\end{tabular}
\label{model params}
\tablefoot{The models used to represent the $L$, $M$, $H$ and $K$ band data taken with MATISSE, GRAVITY and MIRC-X. The parameters are the size (angular diameter or FWHM), the PA of the major axis, the stretch ratio and the flux level. The models used are uniform circular or elliptical disk (CD or ED, respectively), a circular or elliptical Gaussian  (CG or EG, respectively). The stretch ratio is between that of the major and minor axis size. \tablefoottext{a} Flux value was fixed from the start. The size measurements in the $K$ band were obtained using GRAVITY observations only.}

\end{table*}

\end{appendix}

\end{document}